\documentclass[letterpaper]{JHEP3}

\usepackage{amssymb}
\usepackage{amsmath}
\usepackage{bbold}
\usepackage{graphicx}

\def\bea{\begin{eqnarray}}
\def\eea{\end{eqnarray}}
\def\beq{\begin{equation}}
\def\eeq{\end{equation}}

\newcommand{\ignore}[1]{}

\def\<{\langle}
\def\>{\rangle}

\def\Hp{H_{\rm p}}
\def\Hd{H_{\rm b}}
\def\Ap{A_{\rm p}}
\def\Bp{B_{\rm p}}
\def\Ab{A_{\rm b}}
\def\Bb{B_{\rm b}}

\def\u{\overline{u}}
\def\Ne{N_e}

\def\og{\overline{\gamma}}
\def\ob{\overline{\beta}}

\def\vmin{v_{\rm min}}

\def\tobs{\tau_{\rm obs}}

\title{Bubble collisions and measures of the multiverse}

\author{Michael P.~Salem\\
Department of Physics, Stanford University, Stanford, CA 94305}

\abstract{To compute the spectrum of bubble collisions seen by 
an observer in an eternally-inflating multiverse, one must choose a 
measure over the diverging spacetime volume, including choosing an 
``initial'' hypersurface below which there are no bubble 
nucleations.  Previous calculations focused on the case where the 
initial hypersurface is pushed arbitrarily deep into the past.  
Interestingly, the observed spectrum depends on the orientation of 
the initial hypersurface, however one's ability observe the effect 
rapidly decreases with the ratio of inflationary Hubble rates inside 
and outside one's bubble.  We investigate whether this conclusion 
might be avoided under more general circumstances, including 
placing the observer's bubble near the initial hypersurface.  We 
find that it is not.  As a point of reference, a substantial 
appendix reviews relevant aspects of the measure problem of eternal 
inflation.}

\preprint{SU-ITP-11/40}

\begin{document}

\section{Introduction} 
\label{sec:introduction}

String theory argues for the existence of an enormous landscape 
of classically-stable, positive-energy vacua (in addition to 
other states) \cite{Bousso:2000xa,Kachru:2003aw,Susskind:2003kw}.  
The semi-classical methods of Coleman and De Luccia (CDL) indicate 
that such vacua can decay via bubble nucleation, the internal bubble 
geometry described by an open Friedmann--Robertson--Walker (FRW) 
metric, with decay rates (per unit volume) that are generically 
exponentially suppressed \cite{Coleman:1980aw,Lee:1987qc,Brown:2007sd}.  
Meanwhile, in sufficiently long-lived, positive-energy vacua, 
spacetime expands at a rate faster than it succumbs to decay, hence 
the volume grows without bound \cite{Guth:1980zm,Guth:1982pn}.  Thus 
emerges a picture of spacetime in which our local Hubble volume is 
merely part of the inside of an open-FRW bubble, which (likely) 
nucleated in some other bubble, and so on; where in each 
positive-energy bubble other bubbles endlessly nucleate and 
sometimes collide \cite{Garriga:1997ef}.

Although in this picture our local Hubble volume resides entirely
within one such bubble, collisions between our bubble and others 
are potentially observable, if they occur within our past lightcone.  
The probability distribution of observable bubble collisions was 
first estimated by Garriga, Guth, and Vilenkin (GGV) 
\cite{Garriga:2006hw}.  The calculation involves choosing a measure 
over the diverging spacetime volume of potential bubble-nucleation 
sites, including choosing an ``initial'' hypersurface below which 
there are no bubble nucleations.  Motivated in part by the expectation 
of exponentially-suppressed transition rates, GGV took the initial
hypersurface to correspond to a constant-time slice in the infinite 
past (in a spatially-flat de Sitter chart).  Remarkably, the 
spatial distribution of bubble collisions across an observer's sky 
features an anisotropy, indicating the orientation of the initial 
hypersurface, despite its relegation to the infinite past, a phenomenon 
dubbed ``persistence of memory.''

Freivogel, Kleban, Nicolis, and Sigurdson (FKNS) later studied the 
same question in what is expected to be a more realistic cosmology, 
in which the energy density of the vacuum in which our bubble 
nucleates---the ``parent'' vacuum---is much greater than that of the
inflationary epoch within our bubble \cite{Freivogel:2009it}.  In this 
case the effects of the initial hypersurface are heavily suppressed, 
in particular the spatial distribution of bubble collisions becomes 
isotropic except over a solid angle that is too small to reasonably 
expect enough bubble collisions there to reveal the anisotropy.  Thus, 
the relatively small (inflationary) vacuum energy of our bubble 
effectively screens this information about initial conditions.  

While it is reasonable to push the initial hypersurface to past 
infinity, it is worthwhile to consider what might be the signatures 
of other possibilities.  In particular, one might speculate that if 
our bubble nucleates not far from the initial hypersurface, then the 
orientation of the hypersurface might leave some mark in the
spatial distribution of bubble collisions in our sky.  As a point of 
motivation, one might imagine that semi-classical spacetime emerges 
from some more quantum state via what appears within the 
semi-classical spacetime to be a ``tunneling from nothing'' transition 
\cite{Vilenkin:1982de}, with an ``initial,'' near-Planck scale vacuum
rapidly decaying in a cascade of CDL bubble nucleations, one of which
is our bubble.  An immediate objection to this picture is that CDL 
transition rates are proportional to $e^{-B}$, where $B$ is the 
Euclidean action of the instanton minus that of the background (in 
Planck units), and in the semi-classical limit where the 
analysis can be trusted $B$ should be very large.  However, it is 
possible that semi-classical methods give a qualitative description of 
the geometry even at near-Planck scales, in which case transition 
rates may frequently be only weakly suppressed.  Another 
reason one might be skeptical of this possibility is that it assumes 
we live among first wave of bubble nucleations, as opposed to among 
the diverging number of bubbles that nucleate far from the initial 
hypersurface (in the global spacetime, assuming eternal inflation).  
Yet, some phenomenologically viable measures weigh events in spacetime 
according to their occurrence within the vicinity of a single worldline 
originating in the initial vacuum (surveying all semi-classical future 
histories of the worldline).  In this case our residence in one of the 
early bubbles is not necessarily very surprising, since the probability 
of residing in a given bubble is suppressed by the branching ratio 
implied by the series of transitions required to reach it from along a
worldline originating in the initial vacuum.   

In light of possibilities like this, we study the spatial distribution 
of bubble collisions in an observer's sky, while remaining open to the
possibility that our bubble resides near the initial hypersurface below
which there are no bubble nucleations, and being mindful of other 
considerations raised by choosing a measure over the diverging volume
of eternal inflation.  (We include a partial review of the measure 
problem, insofar as it pertains to the issue of bubble collisions, in
the appendix.)  We focus on two basic choices for the
initial hypersurface, (1) the minimal spacelike Cauchy surface in the 
closed de Sitter chart (which can be seen as the surface defining 
initial conditions in the wake of a tunneling-from-nothing event), 
and (2) the null cone representing the past-directed boundary 
of an open de Sitter chart (which can seen to approximate the bubble 
wall of the parent vacuum).  Either of these will look like the GGV 
choice of initial hypersurface in the limit where the hypersurface is 
pushed in the deep past.  The effect of the measure is to prescribe at 
what global time coordinates bubbles like ours typically nucleate (the 
focus of this paper being times on order of the Hubble rate in the 
initial vacuum) and at what FRW radial coordinate observers typically 
arise.

We find that the FKNS prediction of an isotropic distribution of 
bubble collisions is robust with respect to all of these 
considerations; the only scenario in which the distribution is 
anisotropic over an appreciable fraction of the observer's sky is 
when the vacuum energy density of the parent vacuum is not much 
larger than the inflationary energy density in our bubble, in which 
case the likelihood to observe a significant number of bubble 
collisions appears to be small.  (We assume the relevant portion of 
the multiverse has 3+1 large spacetime dimensions.  Significant 
anisotropy in the distribution of bubble collisions can result if 
the parent vacuum has fewer than three large spatial dimensions, 
as in \cite{Salem:2010mi}.)  

This work benefits from techniques largely developed elsewhere 
\cite{Chang:2007eq,Aguirre:2007an,Aguirre:2007wm,Aguirre:2008wy,
Dahlen:2008rd,Aguirre:2009ug}, including the references above.  
Although we do not discuss any specific observational signatures of 
bubble collisions---focusing instead on the potential to observe 
such effects by studying the overlap of causal domains---there is 
a growing body of literature exploring these signals 
\cite{Chang:2008gj,Larjo:2009mt,Czech:2010rg}.  Indeed, 
\cite{Feeney:2010jj,Feeney:2010dd} have developed a search algorithm
for some of these effects in the cosmic microwave background data, 
indicating features consistent with (but not demonstrative of) 
bubble collisions.  For a recent review of these and other topics, 
see \cite{Kleban:2011pg}.

The remainder of this paper is organized as follows.  We establish
the geometry under consideration in Section \ref{sec:geometry}; the
main deviation from previous work occurs in Section 
\ref{sec:hypersurface}, where we describe our choices for the 
initial hypersurface.  The distribution of (potentially) observable 
bubble collisions is calculated in Section \ref{sec:distribution}.  
In Section \ref{sec:V1}, we perform a quick study of the impact of 
other bubble collisions in our past lightcone.  Concluding remarks 
are given in Section \ref{sec:conclusions}.  The appendix provides 
a partial review of the measure problem of eternal inflation, with 
the focus on issues relevant to the study of bubble collisions.

\section{Geometry}
\label{sec:geometry}

We consider cosmologies in which our local Hubble volume is contained 
within an open FRW bubble, ``our bubble,'' which formed via CDL barrier
penetration in a landscape of metastable vacua.  The progenitor of our 
bubble, the ``parent'' vacuum in which our bubble nucleates, has 
positive vacuum energy density, so as to permit such a transition.  We 
take the bubble walls to be thin and approximate their trajectories 
as following null cones emanating from the centers of essentially 
point-like bubble nucleations.  Roughly speaking, these approximations
are valid when the tension of the bubble wall is small compared to the
energy difference between the vacua, in units of the curvature radius
of the parent vacuum \cite{Coleman:1980aw}.  Outside the context of a 
specific model of the landscape, it is unclear how easily this condition 
is made consistent with rapid bubble nucleation rates.  Therefore, our 
calculations should be interpreted as providing only qualitative insight.  
We everywhere work in 3+1 spacetime dimensions.

\subsection{Coordinate systems}
\label{ssec:coords}

Among the consequences of the above approximations is the geometry of 
the parent vacuum is subset of de Sitter (dS) space.  We make use of 
the closed dS chart, with the line element
\beq
ds^2 = \Hp^{-2}\sec^2(T)
\left[-dT^2+dR^2+\sin^2(R)\,d\Omega^2 \right] ,
\eeq
where $-\pi/2<T<\pi/2$ and $0\leq R < \pi$.  Here and below 
$d\Omega^2=d\theta^2+\sin^2(\theta)\,d\phi^2$ gives the line element 
on the unit two-sphere, and $\Hp^{-1}$ denotes the dS curvature 
radius.  It is convenient to refer to an embedding of the geometry in
(4+1)-dimensional Minkowksi space, with
\beq 
ds^2 = -dX_0^2 + dX_1^2 + dX_2^2 + dX_3^2 + dX_4^2 \,.
\eeq
dS space (with curvature radius $\Hp^{-1}$) is then the geometry 
induced on the hyperboloid
\beq
-X_0^2 + X_1^2 + X_2^2 + X_3^2 + X_4^2 = \Hp^{-2} \,.
\label{hyperboloid}
\eeq  
The closed dS chart covers the entire hyperboloid, as indicated
by the embedding
\bea
X_0 \!&=&\! \Hp^{-1}\tan(T) \label{embed2a}\\
X_1 \!&=&\! \Hp^{-1}\sec(T)\sin(R)\cos(\theta) \\
X_2 \!&=&\! \Hp^{-1}\sec(T)\sin(R)\sin(\theta)\cos(\phi) \\
X_3 \!&=&\! \Hp^{-1}\sec(T)\sin(R)\sin(\theta)\sin(\phi) \\
X_4 \!&=&\! \Hp^{-1}\sec(T)\cos(R) \,. \label{embed2e}
\eea

We also make use of the spatially-flat dS chart, with the line 
element 
\beq
ds^2 = -dt^2 + e^{2\Hp t} \left(dr^2 + r^2\,d\Omega^2\right) \,,
\eeq
where $-\infty < t < \infty$ and $0 \leq r < \infty$.  This covers 
only half of the full dS hyperboloid (corresponding to $X_0+X_4>0$), 
as can be seen from the embedding in Minkowski space,
\bea
X_0 \!&=&\! \Hp^{-1}\sinh(\Hp t)+\frac{1}{2}\Hp\, r^2\, e^{\Hp t}
\label{embed1a} \\
X_1 \!&=&\! r\, e^{\Hp t}\cos(\theta) \\
X_2 \!&=&\! r\, e^{\Hp t}\sin(\theta)\cos(\phi) \phantom{\frac{1}{1}}\\
X_3 \!&=&\! r\, e^{\Hp t}\sin(\theta)\sin(\phi) \\
X_4 \!&=&\! \Hp^{-1}\cosh(\Hp t)-\frac{1}{2}\Hp\, r^2\, e^{\Hp t} \,.
\label{embed1e}
\eea
The embedding space makes clear how to transform between the charts.  
Coordinates on the unit two-spheres can be matched trivially.  The 
remaining coordinates are then related by
\beq
e^{-\Hp t} = \frac{\cos(T)}{\sin(T)+\cos(R)} \,, \qquad
\Hp r = \frac{\sin(R)}{\sin(T)+\cos(R)} \,. \label{transform1}
\eeq

Let us now consider our bubble.  The discussion is simpler if we first 
imagine that the geometry of our bubble is also a subset of the dS 
geometry, with curvature radius $\Hd^{-1}$ (with $\Hd\leq\Hp$).  The 
corresponding hyperboloid in the embedding space can be shifted in the 
$X_4$ direction, so as to allow for a continuous (though not smooth) 
matching between it and the parent-vacuum hyperboloid, at what will be 
understood as the bubble wall.  In particular, we momentarily consider 
the bubble geometry to be that induced on the hyperboloid 
\beq
-X_0^2 + X_1^2 + X_2^2 + X_3^2 + (X_4-\Delta_4)^2 = \Hd^{-2} \,.
\label{hyperboloid2}
\eeq
It can be seen that the two hyperboloids intersect on the plane 
$X_4=(\Delta_4^2+\Hp^{-2}-\Hd^{-2})/2\Delta_4$.  This corresponds to
bubble nucleation at $(X_0,X_4)=(0,\Hp^{-1})$, i.e.~$(T,R)=(0,0)$ in 
the closed dS chart (we often suppress the coordinates on the
unit two-sphere).  Per our assumption of negligible initial bubble 
radius, the bubble wall is defined by $X_4=\Hp^{-1}$ (i.e.~the
intersection of the future lightcone of the point of nucleation
and the hyperboloid), which gives
\beq
\Delta_4 = \Hp^{-1} - \Hd^{-1} \,.
\eeq 

Of course, we are interested in bubbles featuring an inflationary and
big-bang cosmology consistent with our observations, as opposed to 
the empty dS space on (\ref{hyperboloid2}).  The geometry of such 
bubbles can be covered with open FRW coordinates, with the line 
element    
\beq
ds^2=-d\tau^2+a^2(\tau)\!\left[d\xi^2+\sinh^2(\xi)\,
d\Omega^2\right] ,
\label{openFRW}
\eeq
where $0\leq \tau < \infty$ (given positive vacuum energy in our bubble)
and $0\leq \xi < \infty$.  It is often convenient to refer to the conformal
time in the bubble, defined by
\beq
\eta = \int \frac{d\tau}{a(\tau)} \,.
\eeq    
The time dependence of the scale factor $a$, and likewise the maximum 
time $\eta_{\rm max}$, depends on the matter content of the 
bubble.  We adopt a simple approximation of the standard inflationary 
plus big-bang cosmology, the details of which are described as 
they become relevant below.  It is possible to embed this generic open 
FRW geometry in the higher-dimensional Minkowski space.  With some 
hindsight regarding the appearance of $\Delta_4$, we write
\bea
X_0 \!&=&\! a(\tau)\,\cosh(\xi) \label{embed3a}\\
X_1 \!&=&\! a(\tau)\,\sinh(\xi)\cos(\theta) \\
X_2 \!&=&\! a(\tau)\,\sinh(\xi)\sin(\theta)\cos(\phi) \\
X_3 \!&=&\! a(\tau)\,\sinh(\xi)\sin(\theta)\sin(\phi) \\
X_4 \!&=&\! f(\tau) + \Delta_4 \,, 
\label{embed3e}
\eea
where the function $f(\tau)$ corresponds to a solution of the 
differential equation
\beq
\dot{f}^2= \dot{a}^2-1 \,,
\eeq  
dots here denoting derivatives with respect to $\tau$.  

Although the CDL instanton generates a homogeneous and isotropic FRW 
bubble, it is necessary for there to be a round of slow-roll inflation 
within the bubble.  This is to redshift away the large initial spatial 
curvature of the bubble, so as to agree with our observation of an 
approximately flat FRW geometry.  For simplicity, we treat the inflaton 
energy density as equivalent to vacuum energy density, taking 
$\rho_{\rm inf} = 3\Hd^2/8\pi G$.  Furthermore we assume nothing 
else contributes significantly toward the energy density of the bubble, 
until reheating.  The scale-factor solution before reheating is then 
\beq
a(\tau) = \Hd^{-1}\sinh(\Hd\tau) \,, \qquad
a(\eta) = \Hd^{-1}{\rm csch}(-\eta) \,,
\label{sf}
\eeq
which incorporates the appropriate CDL boundary conditions.  We have
set an integration constant so that proper and conformal time in the 
bubble are related during inflation by
\beq
\eta = \ln\!\big[\!\tanh(2\Hd\tau)\big] \,.
\eeq   
This scale-factor solution corresponds to the open dS chart, and is 
embedded into the higher-dimensional Minkowski space by inserting it
into (\ref{embed3a})--(\ref{embed3e}), with 
$f(\tau)=\Hd^{-1}\cosh(\Hd\tau)$.  This inflationary geometry therefore
coincides with the dS geometry induced on the hyperboloid 
(\ref{hyperboloid2}), and therefore matches continuously onto the parent 
vacuum geometry at the bubble wall.  Of course, the geometry of the bubble 
after reheating is no longer locally dS space, and will therefore deviate 
from the dS hyperboloid.

\subsection{The initial hypersurface}
\label{sec:hypersurface}

Our cosmological setup involves an initial hypersurface, below which 
we presume there are effectively no bubble nucleations.  We consider two 
possibilities, which are most easily described starting with pure dS space 
(i.e.~without any bubble nucleations).  Then, the hypersurface $T=0$ could 
correspond to the surface of initial conditions in a ``tunneling from 
nothing'' scenario \cite{Vilenkin:1982de}, while the hypersurface $T=R$ 
could correspond to what might roughly appear like an initial hypersurface 
if for example the bubble wall of the parent vacuum is such that it screens 
the effects of all bubble nucleations outside of it.  For brevity we 
henceforth refer to these as the ``tunneling from nothing'' and ``bubble 
wall'' initial hypersurfaces.  

These hypersurfaces break a dS symmetry, a consequence of which is we 
cannot place the nucleation of our bubble at $T=0$ without loss 
of generality.  (The remaining dS symmetries still allow us to place 
the nucleation of our bubble at $R=0$.)  Instead, our bubble 
generically nucleates at some point $(T,R)=(T_{\rm n},0)$, $T_{\rm n}\geq 0$.  
Note however that the point $(T,R)=(T_{\rm n},0)$ can be translated to the 
point $(T,R)=(0,0)$ by performing a boost in the embedding Minkowski
space; in particular by taking
\bea
X'_0 \!&=&\! \gamma (X_0 - \beta X_4) \\
X'_4 \!&=&\! \gamma (X_4 - \beta X_0) \label{boostT2} \,,
\eea     
where $X_1$, $X_2$, and $X_3$ are unchanged, and where 
$\gamma=(1-\beta^2)^{-1/2}$ with 
\beq
\gamma=\sec(T_{\rm n}) =\cosh(\Hp\tau_{\rm n}/2)\,, 
\qquad {\rm and} \qquad
\beta = \sin(T_{\rm n}) = \tanh(\Hp\tau_{\rm n}/2) \,.
\label{betadef}
\eeq
Here $\tau_{\rm n}$ is the proper time between $T=0$ and $T=T_{\rm n}$,
keeping all other coordinates fixed at the origin.  Thus, we have the 
option of working with the initial hypersurfaces $T=0$ or $T=R$, with 
our bubble nucleating at $T=T_{\rm n}$, or working with the boosted 
initial hypersurfaces (see below), with our bubble nucleating at $T=0$.  
We choose the latter.  

\begin{figure*}[t!]
\begin{center}
\begin{tabular}{ccc}
\includegraphics[width=0.33\textwidth]{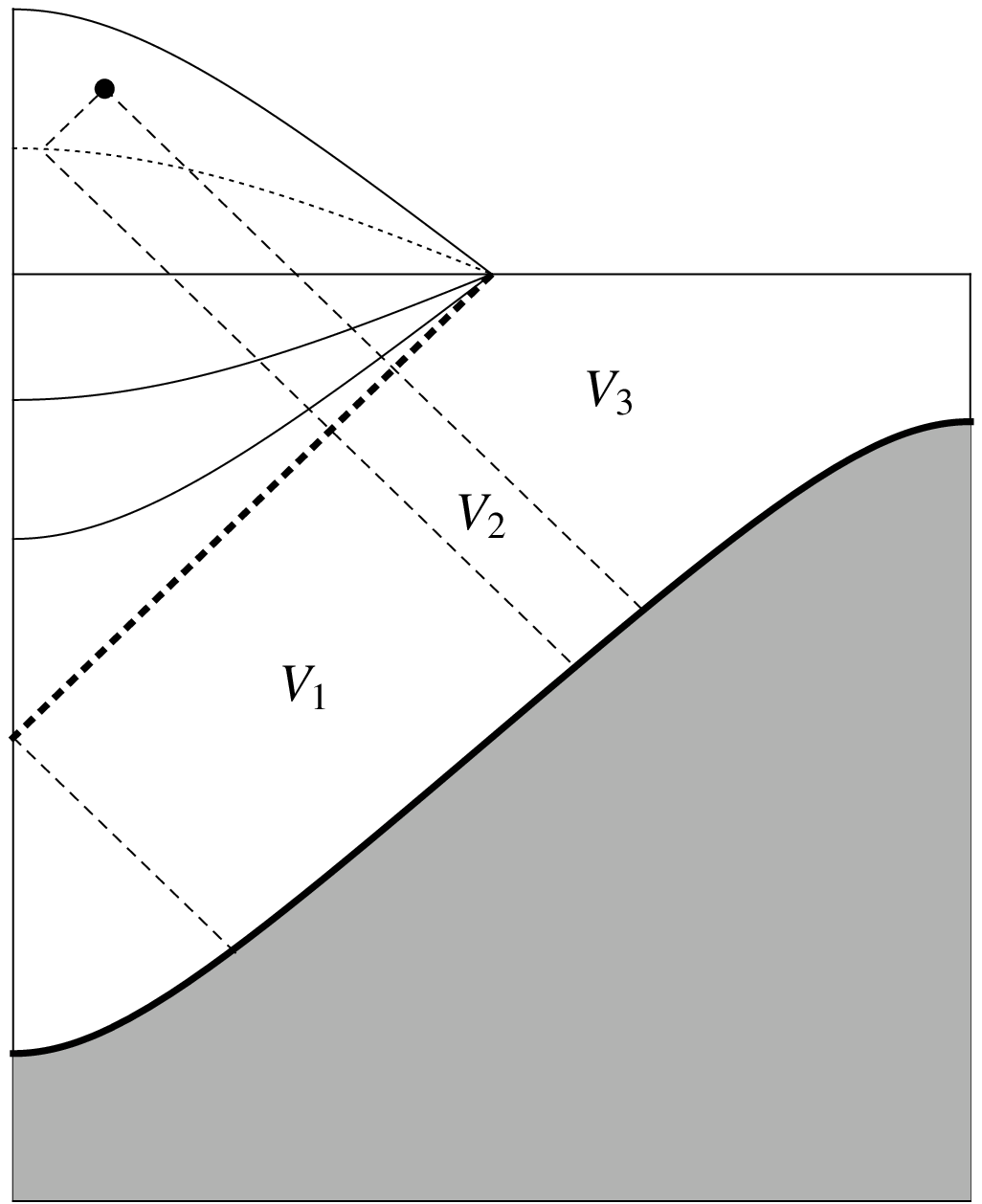} 
& \phantom{spacespace} &
\includegraphics[width=0.33\textwidth]{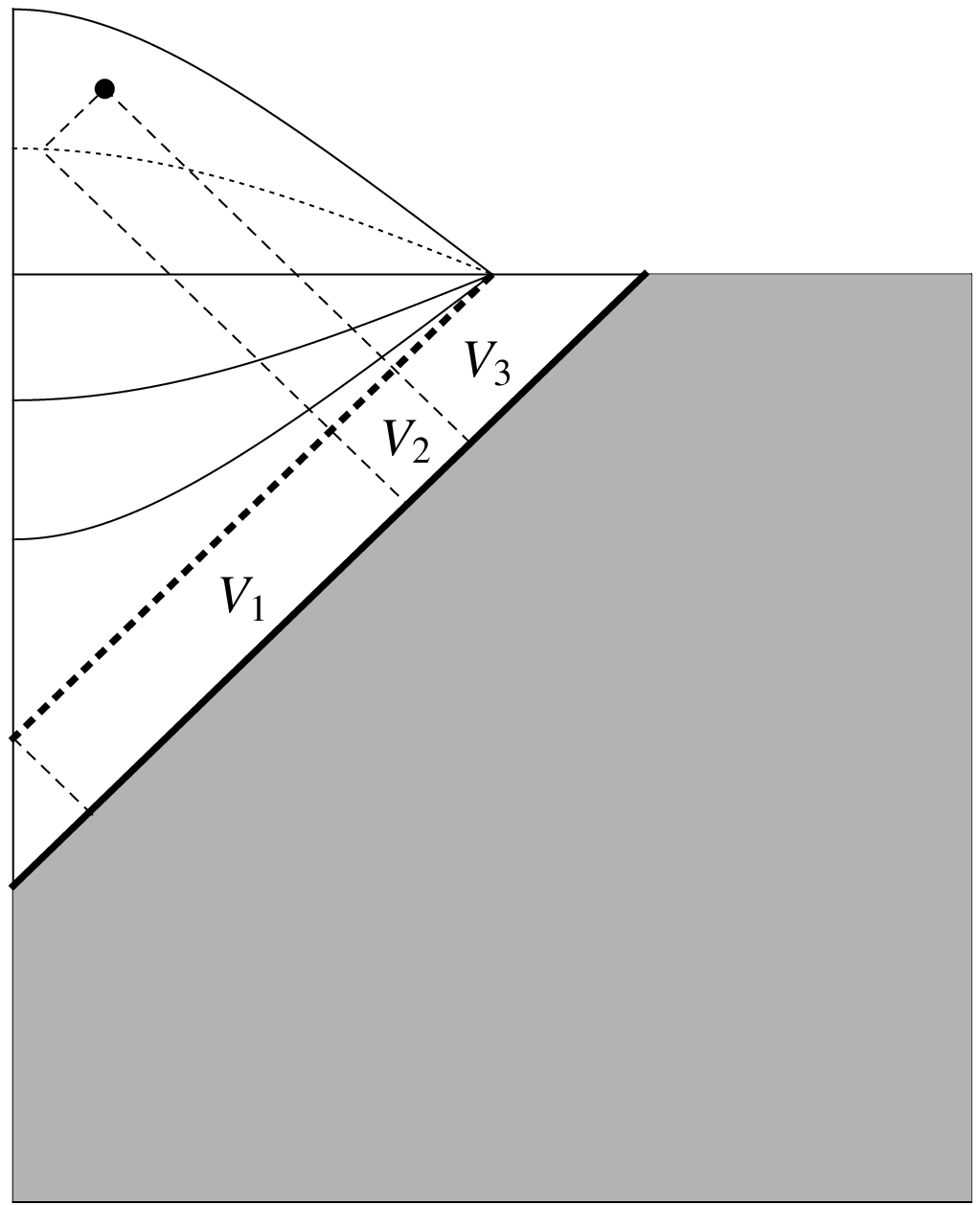} 
\end{tabular}
\caption{\label{fig:C} Toy conformal diagrams of our bubble nucleating 
in the parent vacuum, suppressing the unit two-sphere.  The bubble wall 
corresponds to the thick, dotted line, while the initial hypersurfaces 
(2.31) and (2.32) correspond to the thick, solid lines in the left and 
right diagrams, respectively.  For future reference the dotted line 
represents the time of recombination, the large dot an observer, and 
the dashed lines delineate the volume in which the future lightcone of 
an event intersects but does not entirely cover the surface of last 
scattering.}
\end{center}
\end{figure*}

The would-be initial hypersurface $T=0$ corresponds to what would have 
been the plane $X_0=0$ in the embedding coordinates.  Boosting this 
hypersurface so as to translate our bubble to the origin, we have
$X_0\to\gamma(X'_0+\beta X'_4) = 0$, which corresponds to
\beq
\sin(T) = -\sin(T_{\rm n})\cos(R) \,.
\label{h1}
\eeq 
Meanwhile, the would-be hypersurface $T=R$ corresponds to what would
have been the plane $X_4=\Hp^{-1}$ in the embedding coordinates.  
Boosting this hypersurface so as to translate our bubble to the origin, 
we have $X_4\to\gamma(X'_4+\beta X'_0)=\Hp^{-1}$, 
which corresponds to
\bea
T=R-T_{\rm n} \,.
\label{h2}
\eea
Two toy conformal diagrams of the bubble plus parent vacuum geometry, 
including the boosted initial hypersurfaces, are displayed in Figure 
\ref{fig:C}.

It is convenient to perform another boost, so as to translate an 
observer in our bubble from the random position  
$(\xi,\,\theta,\,\phi)=(\xi_0,\,0,\,0)$, to the more central position
position $(0,\,0,\,0)$.  (Rotational invariance has allowed us to set 
$\theta_0=\phi_0=0$ without loss of generality.  Although unbroken 
FRW and dS symmetries would also allow us to set $\xi_0=0$, we have 
already exploited the corresponding symmetry to set $R_{\rm n}=0$.)  
In light of the embedding (\ref{embed3a})--(\ref{embed3e}), we see 
this can be accomplished (for arbitrary scale-factor $a$) with the 
boost
\bea
X'_0 \!&=&\! \og(X_0-\ob X_1) \label{boost1}\\
X'_1 \!&=&\! \og(X_1-\ob X_0) \,, \label{boost2}
\eea 
keeping $X_2$, $X_3$, and $X_4$ fixed, where 
$\og=(1-\ob \!\!\!\phantom{\beta}^2)^{-1/2}$ with
\beq
\og=\cosh(\xi_0)\, \qquad {\rm and} \qquad  
\ob=\tanh(\xi_0) \,.
\eeq
Note that this boost leaves intact the trajectory of the bubble wall: 
in terms of the embedding coordinates, we have $X'_4=X_4=\Hp^{-1}$.  

Aside from translating the observer to the ``center'' of our bubble, 
the effect of the above boost is to perform a second rotation on the
initial hypersurface.  For the initial hypersurface of (\ref{h1}), we 
have $\gamma(X_0+\beta X_4)\to\gamma[\og(X'_0+\ob X'_1)+\beta X'_4]=0$, 
or
\beq
\og\sin(T)+\beta\cos(R)+\og\ob\sin(R)\cos(\theta)=0 \,.
\label{h1b}
\eeq
This describes the embedding of the ``tunneling from nothing'' initial
hypersurface as seen by an observer in our bubble who places herself
at $\xi=\theta=\phi=0$.  
For the initial hypersurface (\ref{h2}), we have
$\gamma(X_4+\beta X_0)\to \gamma[X'_4+\og\beta(X'_0+\ob X'_1)]=\Hp^{-1}$,
which corresponds to 
\beq
\gamma\og\beta\sin(T)+\gamma\cos(R)
+\gamma\og\beta\ob\sin(R)\cos(\theta)=\cos(T) \,.
\label{h2b}
\eeq
This describes the embedding of the ``bubble wall'' initial 
hypersurface as seen by the above-mentioned observer.
Note that this boost has introduced $\theta$-dependence in both 
initial hypersurfaces.  This can be seen in Figure \ref{fig:C} in
the following way.  Take the toy conformal diagrams there to 
display the geometry for $\theta=0$ ($\phi=0$).  Then we can 
augment the diagrams with the geometry at $\theta=\pi$ ($\phi=0$)
by attaching the mirror image of each diagram to its left.  Then 
it can be seen that the observer, by virtue of residing to the 
right of the origin, has more of the initial hypersurface in her 
past lightcone to the right ($\theta=0$) than to the left 
($\theta=\pi$).      

The ultimate cause of this anisotropy is the breaking of a 
background dS symmetry by the choice of initial hypersurface, and it 
is the root of the ``persistence of memory'' effect mentioned in a 
slightly different context in the introduction.  In particular, if 
we take the late-nucleation-time limit $T_{\rm n}\to \pi/2$ 
(corresponding to $\gamma\to\infty$ and $\beta\to 1$), we find both 
initial hypersurfaces (\ref{h1b}) and (\ref{h2b}) become 
\beq
\og\sin(T)+\cos(R)+\og\ob\sin(R)\cos(\theta)=0 \,,
\eeq
which corresponds to the GGV surface mentioned in the introduction.

\subsection{Matching coordinates inside and outside the bubble}
\label{sec:matching}

Our calculations require us to map null rays from the parent vacuum 
into our bubble.  We are only interested in incoming, radial null rays, 
and we match them by identifying coordinates at the bubble wall, using 
the Minkowski embedding space described in Section \ref{ssec:coords}.      
In terms of the closed dS chart of the parent vacuum, we can label a 
given null ray according to the coordinates $(T_{\rm c},R_{\rm c})$ at 
which it originates, its trajectory given by $T=-R+T_{\rm c}+R_{\rm c}$.
In this chart the bubble wall corresponds to $T=R$, and so according
to the embedding (\ref{embed2a}) we have 
$X_0 = \Hp^{-1}\tan\left[\frac{1}{2}(T_{\rm c}+R_{\rm c})\right]$
at the bubble wall.  In the open dS chart of our bubble, the bubble
wall corresponds to $\eta\to-\infty$ and $\xi\to\infty$, with $\eta+\xi$ 
held constant, and so according to the embedding 
(\ref{embed3a}) with the scale-factor solution (\ref{sf}) we have
$X_0 = \Hd^{-1}e^{\eta+\xi}$.  Performing the matching, we find 
its trajectory in terms of the FRW bubble coordinates to be
\beq
\eta = \ln\!\Big\{(\Hd/\Hp)
\tan\!\left[\textstyle\frac{1}{2}(T_{\rm c}+R_{\rm c})\right]\!\Big\} 
-\xi \,.
\label{matching}
\eeq

For later reference we also construct a mapping between certain 
geodesics in the parent vacuum and their extensions into our bubble.  
The calculation is a bit technical, and benefits from reference to
\cite{Vilenkin:1996ar,Harlow:2011az} (see \cite{Israel:1966rt,Blau:1986cw} 
for more background).  Again, we make use of the Minkowski embedding
space, and suppress the unit two-sphere by defining
\beq
X \equiv\sqrt{X_1^2+X_2^2+X_3^2} \,.
\eeq  
Our approach takes advantage of the fact that geodesics in dS 
correspond to the intersection of the dS hyperboloid and plane 
hypersurfaces that pass through the origin of the hyperboloid, in the
embedding geometry.\footnote{The author thanks Daniel Harlow and 
Leonard Susskind for pointing this out.}  
For geodesics in the parent vacuum, we write the plane
\beq
\Ap X_0 - X + \Bp X_4 = 0 \,.
\label{geop}
\eeq  
For example, a geodesic with constant radial coordinate $r$ in the 
spatially-flat dS chart has $\Ap=\Bp=\Hp r$, as can be seen via the 
embedding (\ref{embed1a})--(\ref{embed1e}).  A geodesic with constant 
radial coordinate $R$ in the closed dS chart has $\Ap=0$, $\Bp=\tan(R)$,
c.f.~(\ref{embed2a})--(\ref{embed2e}).  We can use (\ref{boostT2}) to 
boost this plane so as to describe a geodesic that is orthogonal to the 
initial hypersurface (\ref{h1b}); this gives
$\Ap=\tan(T_{\rm n})\tan(R)$ and $\Bp=\sec(T_{\rm n})\tan(R)$.

Since we approximate the inflationary phase within our bubble as pure 
vacuum-energy domination, geodesics in the inflating spacetime can also 
be described as the intersection of a plane hypersurface and a dS 
hyperboloid.  In this case we write the plane     
\beq
\Ab X_0 - X + \Bb (X_4-\Delta_4) = 0 \,,
\label{geob}
\eeq
where $\Delta_4$ accounts for the translated origin of the dS 
hyperboloid describing the inflationary epoch within our bubble,
(\ref{hyperboloid2}).  In order for this geodesic to describe the 
extension of the parent-vacuum geodesic given by (\ref{geop}), the 
plane (\ref{geob}) should intersect (\ref{geop}) at the bubble wall 
(given by $X_4=\Hp^{-1}$), and the tangents to the geodesics should each
make the same angle with the tangent to the bubble wall (at the point of 
intersection).  Solving for $\Ab$ and $\Bb$ in terms of $\Ap$ and $\Bp$ 
in this way provides a map between geodesics in the parent vacuum and 
geodesics in the inflating region of the bubble.  Since during inflation 
in the bubble geodesics rapidly asymptote to comoving in the bubble FRW 
frame, this serves as a map between geodesics in the parent vacuum and 
comoving geodesics in the bubble.

The algebra is a bit messy, so we only lay it out symbolically.  
Combining (\ref{geop}) with the hyperboloid constraint (\ref{hyperboloid}), 
we can solve for the components $X$ and $X_4$ of the geodesic as a function 
of $X_0$.  We thus write the geodesic as a curve in the embedding space, 
\beq
g^\mu_{\rm p}(X_0) = \{ X_0,\, X(X_0),\, X_4(X_0) \} \,,
\eeq 
where the index $\mu$ is understood to run over the three components of 
the vector.  Likewise, combining (\ref{geob}) with (\ref{hyperboloid2}), 
we obtain the curve $g^\mu_{\rm b}(X_0)$.  The curve describing the 
bubble wall is simply $g^\mu_{\rm w}(X_0)=\{X_0,\,X_0,\,\Hp^{-1}\}$.  
Using this we can compute the (Minkowski) times at which the curves 
$g^\mu_{\rm p}$ and $g^\mu_{\rm b}$ intersect the bubble wall.  These 
are, respectively 
\beq
X_0 = \frac{\Bp}{\Hp(1-\Ap)}\,, \qquad X_0 = \frac{\Bb}{\Hd(1-\Ab)} \,.
\label{Tatbubble}
\eeq
We denote the tangents to $g^\mu_{\rm p}$ and $g^\mu_{\rm b}$ with 
$t^\mu_{\rm p}$ and $t^\mu_{\rm b}$, respectively.  For example,
\beq
t^\mu_{\rm p}(X_0) = \frac{dX_0}{ds}\frac{dg^\mu_{\rm p}}{dX_0}
=\left(-1+\frac{dX^2}{dX_0^2}+\frac{dX_4^2}{dX_0^2}\right)^{\!\!-1/2}\!
\frac{dg^\mu_{\rm p}}{dX_0} \,,
\eeq  
where $ds$ is an infinitesimal proper time interval, 
$ds^2 = \eta_{\mu\nu}dX^\mu dX^\nu$, with $\eta_{\mu\nu}$ denoting
the Minkowski metric on the embedding space with the unit two-sphere
suppressed.  For the tangent to the bubble wall we can simply take 
$t^\mu_{\rm w}=\{1,\,1,\,0\}$.  As described above, we solve for $\Ab$ 
and $\Bb$ in terms of $\Ap$ and $\Bp$ by solving
\beq
g^\mu_{\rm p} = g^\mu_{\rm d} \qquad {\rm and} \qquad
\eta_{\mu\nu}t^\nu_{\rm w}t^\mu_{\rm p} 
= \eta_{\mu\nu}t^\nu_{\rm w}t^\mu_{\rm b} \,,
\eeq
where all expressions are evaluated at the bubble wall, i.e.~at 
(\ref{Tatbubble}).  The solution is
\beq
\Ab = \frac{2\Hp\Hd\Bp(1-\Ap)}{2\Hp^2(1-\Ap)+(\Hp^2-\Hd^2)\Bp^2} \,,\qquad
\Bb = \frac{2\Hp^2\Ap(1-\Ap)+(\Hp^2-\Hd^2)\Bp^2}
{2\Hp^2(1-\Ap)+(\Hp^2-\Hd^2)\Bp^2} \,. \,
\label{AB}
\eeq  

The inflationary geometry within the bubble is embedded by using 
(\ref{embed3a})--(\ref{embed3e}) with the scale-factor solution 
(\ref{sf}).  Inserting this into (\ref{geob}), one can solve 
for the radial coordinate $\xi$ of the geodesic in terms of $\Ab$
and $\Bb$, in the limit $\Hd\tau\gg 1$.  This gives
\beq
\xi = \ln\!\left[\frac{\Hd\Bb+\Hp\sqrt{1-\Ab^2+\Bb^2}}{\Hp(1-\Ab)}\,\right] .
\eeq
For a geodesic that is initially comoving at radial coordinate $r$ 
with respect to the spatially-flat dS chart, combining this with
(\ref{AB}) and the results below (\ref{geop}) gives
\beq
\xi = \ln\!\left(\frac{1+\Hd r}{1-\Hp r}\right) .
\label{r2xi}
\eeq
For a geodesic that is orthogonal to the initial hypersurface 
(\ref{h1b}) in the closed dS chart, 
\beq
\xi = \ln\!\left[\frac{\Hp\cos(T_{\rm n})+\Hd\sin(R)}
{\Hp\cos(R+T_{\rm n})}\right] .
\label{R2xi}
\eeq 
Although the above matching has been performed only with respect to 
the inflationary geometry within the bubble, the dS coordinate $\xi$ 
matches trivially onto the post-inflationary open-FRW coordinate 
denoted by the same symbol.  

Finally, for later reference we also compute the scale-factor time
between some reference hypersurface---here taken to be the surface of 
fixed time $t=0$ in the spatially-flat dS chart---and a fixed FRW 
time $\tau_{\rm ref}\gg\Hd^{-1}$ hypersurface in the bubble.  
Consider a congruence of initially constant-$r$ geodesics that enter 
our bubble and become comoving along the set of radial coordinates 
$\xi$, according to (\ref{r2xi}).  Note that, tracing along these 
geodesics, a small coordinate separation $\Delta r$ at 
$t=0$ evolves into a comoving coordinate separation
\beq
\Delta \xi = \frac{(\Hp+\Hd)\,\Delta r}{(1-\Hp r)(1+\Hd r)} 
= \frac{(\Hp+\Hd e^{-\xi})^2\,e^\xi\,\Delta r}{\Hp+\Hd} \,,
\eeq    
at $\tau=\tau_{\rm ref}$.  Thus, a radial comoving rod covering 
physical distance $\Delta r$ at $t=0$ covers a physical distance 
$a(\tau_{\rm ref})\Delta\xi=(1/2)\Hd^{-1}e^{\Hd\tau_{\rm ref}}\Delta\xi$ 
at $\tau=\tau_{\rm ref}$.  Moreover, a comoving ``annulus'' centered 
at $r=0$ with physical three-volume $4\pi\,r^2 \Delta r$ at $t=0$ 
covers an annulus centered at $\xi=0$ with physical three-volume
$(\pi/2)\Hd^{-3}e^{3\Hd\tau_{\rm ref}}\sinh^2(\xi)\,\Delta\xi$ at
$\tau=\tau_{\rm ref}$.  Defining the scale-factor time to be one 
third the logarithm of the implied expansion factor, we see  
\beq
\Delta t_{\rm sf} = \Hd\tau_{\rm ref} + \xi + 
\frac{1}{3}\ln\!\left[\frac{(1+e^{-\xi})^2(\Hp+\Hd e^{-\xi})^4}
{32\Hd^3(\Hp+\Hd)}\right] ,
\label{sftime}
\eeq
where we have exploited the symmetry of the annulus to equate 
the change in logarithm of its volume expansion factor with the 
change in logarithm of the local volume expansion factor.

\subsection{Bubble collision geometry}
\label{sec:collision}

With the parent vacuum and our bubble in place, we now introduce an 
additional bubble, the colliding bubble, the effects of which we hope 
to observe.  The collision between the colliding bubble and our bubble 
occurs along the bubble walls, but in our approximation of point-like
bubble nucleation with thin bubble walls, this corresponds to at the 
intersection of the future lightcones of the bubble nucleation events.  
We focus on the possibility that the bubble collision leaves some 
imprint on the cosmic microwave background---small enough to have so 
far evaded unambiguous detection, but not so small so as to be 
undetectable---in which case we are interested in the intersection of 
the past lightcone of the observer, the hypersurface of recombination, 
and the future lightcone of the colliding bubble nucleation event.

\begin{figure*}[t!]
\begin{center}
\begin{tabular}{ccc}
\includegraphics[width=0.33\textwidth]{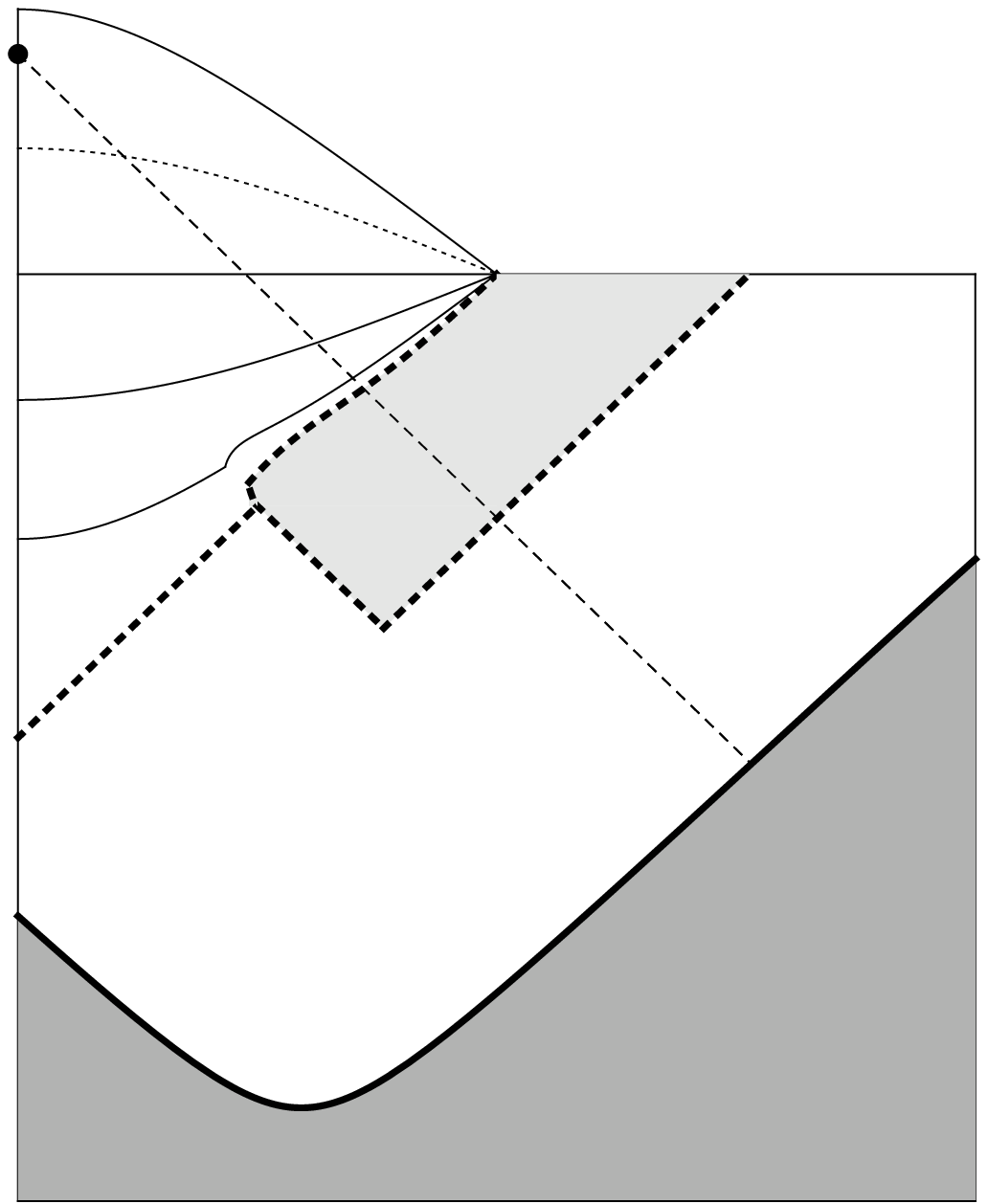} 
& \phantom{spacespace} &
\includegraphics[width=0.33\textwidth]{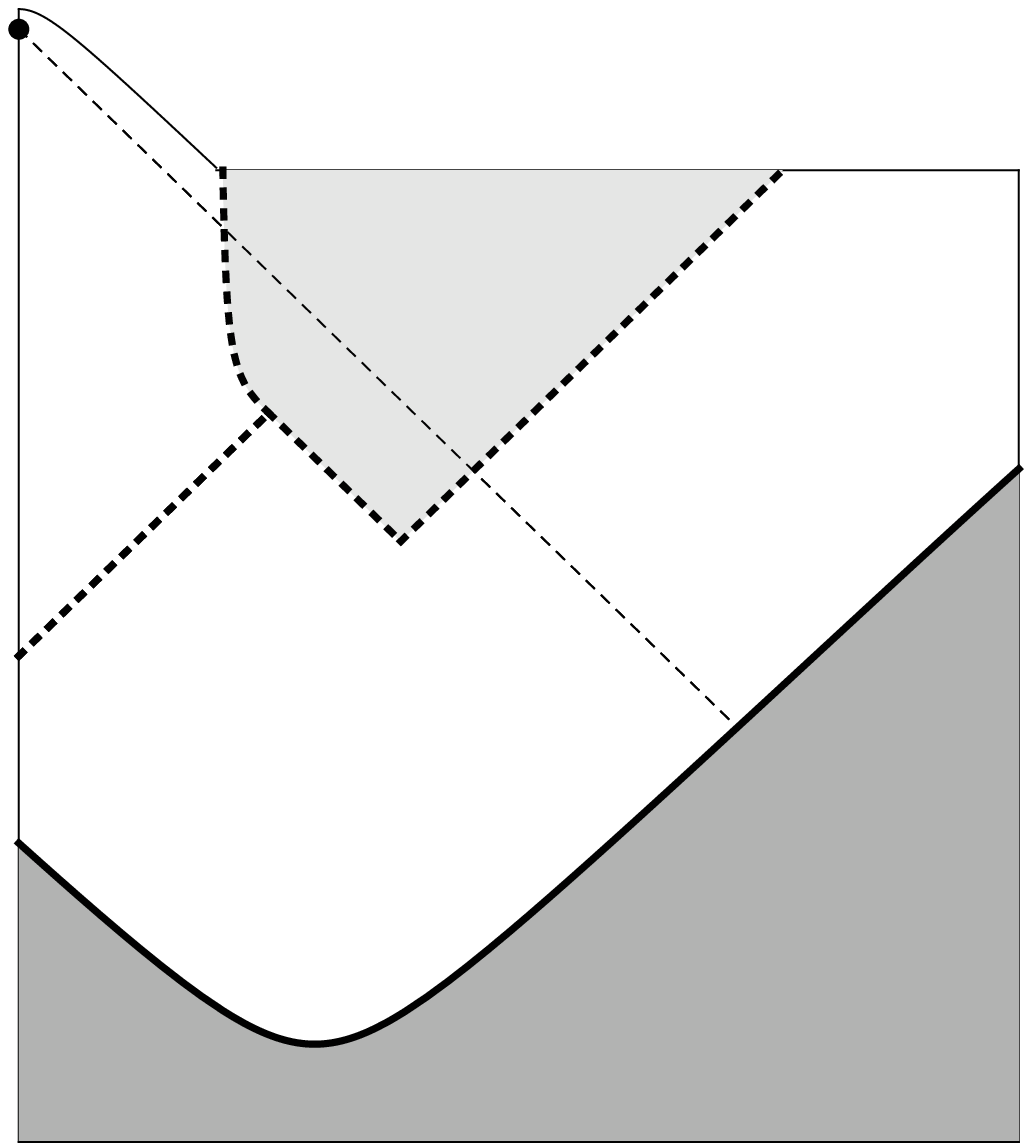} 
\end{tabular}
\caption{\label{fig:C2} Toy conformal diagrams of our bubble colliding 
with another bubble (the colliding bubble is lightly shaded).  The 
bubble/domain walls correspond to the thick, dotted lines, and we 
have chosen the initial hypersurface (2.36) for an observer who has 
been translated to $\xi=0$.  The left panel displays a cartoon of a 
bubble collision that does not significantly perturb the global, open 
FRW geometry in our bubble; the right panel displays a cartoon of one 
that does.}
\end{center}
\end{figure*}

Although our analysis avoids reference to the microphysics of the 
collision, it might help to draw a more concrete picture.  Consider a
landscape effectively described by some number of scalar fields, the 
metastable states in the landscape corresponding to local minima in 
the scalar-field potential.  Bubble nucleations occur via CDL barrier 
penetration, but in vacua like ours the instanton itself does not take 
the tunneling field all the way to the local minimum; instead it 
arrives at rest on the other side of the potential barrier, drives 
slow-roll inflation as it classically evolves down a shallow slope in 
the potential, and reheats the bubble as the field finally oscillates 
around the local minimum.  When our bubble collides with another one, 
a domain wall forms between our vacuum and another one (the other 
vacuum is not necessarily that of the colliding bubble 
\cite{Easther:2009ft,Johnson:2010bn}).  This domain wall moves away 
from the center of our bubble, unless the vacuum energy behind the
domain wall is smaller than ours, in which case the tension of the
domain wall is important for determining its trajectory 
\cite{Chang:2007eq}.  In the first case, the domain wall does not 
follow precisely what would have been the trajectory of the bubble 
wall in the absence of a collision, and so the domain wall can perturb 
the initial value of the inflaton field in the bubble.  (The left 
panel of Figure \ref{fig:C2} provides a cartoon illustration.)  This 
perturbation might be large, but it is redshifted during inflation, 
and if inflation does not last too long (not more than roughly ten 
$e$-folds more than is necessary to solve the flatness problem), it 
might produce small but observable signatures 
\cite{Chang:2008gj,Larjo:2009mt,Czech:2010rg}.

On the other hand, if the aforementioned domain wall moves toward 
the center of our bubble, it significantly disrupts the FRW geometry 
in its wake.  (The right panel of Figure \ref{fig:C2} provides a 
cartoon illustration.)  In regions within our bubble where inflation 
proceeds more or less as before, we have the same story as above.  
In regions where inflation is significantly disrupted (including 
regions in the colliding bubble vacuum on the other side of the 
domain wall), it is reasonable to expect that observers like us who 
reside in large galaxies do not arise to observe the effects.  At 
this point, we must simply assume that most of the volume within our 
bubble (after regulating the diverging volume with a spacetime 
measure, see the appendix) falls into one of these categories, so 
that it is not unusual for observers like us to observe a (nearly) 
flat FRW Hubble volume as we do.

Proceeding now with our analysis, we denote the location of the point 
of nucleation of the colliding bubble  
$(T,\,R,\,\theta,\,\phi)=(T_{\rm c},R_{\rm c},\,\theta_{\rm c},\,0)$, 
where we have used the remaining symmetry on the unit two-sphere to 
place both the observer and the center of the colliding bubble 
nucleation event at $\phi=0$.  For notational simplicity we work on a 
hypersurface of constant $\phi$ and suppress the dependence on $\phi$; 
in the end the complete results are easily obtained by utilizing the 
rotational symmetry of the problem.

It is complicated to work out the coordinates of the future lightcone 
of the colliding bubble nucleation event, let alone transform and 
evolve them in terms of our bubble coordinates.  Instead we approximate 
the relevant portion of the lightcone---that is, the portion that will
intersect the past lightcone of the observer and the surface of last
scattering---as a constant ``front'' (following 
\cite{Freivogel:2009it}).  That is, we start with the future lightcone 
in the $\theta=\theta_{\rm c}$ plane,
\beq
R=R_{\rm c} \pm (T-T_{\rm c}) \,.
\eeq
Using (\ref{matching}), we match the null ray onto the open 
FRW coordinates of our bubble, giving
\beq
\xi = \ln\!\Big\{(\Hd/\Hp)
\tan\!\left[\textstyle\frac{1}{2}(T_{\rm c}+R_{\rm c})\right]\!\Big\} 
-\eta \,,
\label{xiandeta}
\eeq
The observable portion of the future lightcone of the colliding 
bubble nucleation event is then approximated as a plane orthogonal 
to this trajectory, i.e.
\beq
\xi= \sec(\theta-\theta_{\rm c})\! \left[ \ln\!\Big\{(\Hd/\Hp)
\tan\!\left[\textstyle\frac{1}{2}(T_{\rm c}+R_{\rm c})\right]\!\Big\} 
-\eta \right] \,.
\eeq
The above approximation is accurate because the intrinsic 
curvature on the hypersurface corresponding to the future 
lightcone of the colliding bubble, after propagation into our 
bubble, is of order the curvature radius of our bubble, which 
is empirically much larger than the distance to (and across) the 
surface of last scattering.

In Section \ref{sec:hypersurface} we performed a boost in the 
Minkowski embedding geometry, so as to place an arbitrary observer 
in the bubble at $\xi=0$.  The surface of last scattering for this 
observer then corresponds to the two-sphere with radial coordinate 
$\xi_\star=\eta_0-\eta_\star$, on the hypersurface $\eta=\eta_\star$, 
where $\eta_0$ is the FRW (conformal) time of the observer and 
$\eta_\star$ is the time at recombination.  The intersection of the 
surface of last scattering and the future lightcone of the colliding 
bubble nucleation event then corresponds to the region 
$\xi(\eta_\star,\,\theta)\leq\xi_\star$.  Thus, the effects of
the bubble collision are limited to 
\beq
\xi_\star\cos(\theta-\theta_{\rm c})\geq \ln\!\Big\{(\Hd/\Hp)
\tan\!\left[\textstyle\frac{1}{2}(T_{\rm c}+R_{\rm c})\right]\!\Big\}
- \eta_\star \,.
\label{m&m}
\eeq

Let us discuss the meaning of (\ref{m&m}).  When 
$(\Hd/\Hp)\tan\!\left[\frac{1}{2}(R_{\rm c}+T_{\rm c})\right]
> e^{\eta_\star+\xi_\star}$, the collision occurs too late for its 
future lightcone to intersect the surface of last scattering.  
These bubble collisions are unobservable by virtue of causality.  
They correspond to bubble nucleations in the region marked $V_3$ 
in Figure \ref{fig:C} (note that Figure \ref{fig:C} displays the 
geometry before the boost that takes $\xi_0\to0$).  When 
$(\Hd/\Hp)\tan\!\left[\frac{1}{2}(R_{\rm c}+T_{\rm c})\right]
>e^{\eta_\star-\xi_\star}$, the collision occurs so early that its 
future lightcone covers the entire cosmic microwave sky of the 
observer.  These bubble collisions 
are in principle observable, but the redshifting of their features
during inflation combined with the lack of contrast with an 
unaffected region on the cosmic microwave background would seem to 
make them undetectable.  They correspond to bubble nucleations in 
the volume marked $V_1$ in Figure \ref{fig:C}.  Only bubble 
nucleations in the volume marked $V_2$ in Figure \ref{fig:C} 
intersect the surface of last scattering so as give a real, 
non-zero solution for $\theta$ in (\ref{m&m}), corresponding to 
the angular scale of the collision region on the observer's sky.  
We consider only these bubble collisions potentially observable.

Before proceeding, for future reference we compute the solid angle
subtended by the intersection of our bubble wall and the colliding
bubble wall, as a function of time.  That is, at any given time 
the bubble wall of our bubble is a two-sphere, as is the bubble wall
of any colliding bubble, and we compute the solid angle of the 
intersection of these two two-spheres, from the perspective of 
within our bubble.  As always we suppress the $\phi$ coordinate, 
and for simplicity we take the colliding bubble to nucleate at
$\theta_{\rm c}=0$ (symmetry guarantees that the final result is
independent of $\theta_{\rm c}$).  It is easiest to work in terms
of the Minkowski embedding coordinates.  Then the future lightcone 
of the colliding bubble corresponds to the intersection of the 
future lightcone of its nucleation point,
\beq
\left[X_1-\frac{1}{\Hp}\frac{\sin(R_{\rm c})}{\cos(T_{\rm c})}\right]^2 
+ X_2^2 + X_3^2 
+ \left[X_4-\frac{1}{\Hp}\frac{\cos(R_{\rm c})}{\cos(T_{\rm c})}\right]^2 
= \big[X_0-\Hp^{-1}\tan(T_{\rm c})\big]^2 \,,
\eeq
and the dS hyperboloid of the parent vacuum, (\ref{hyperboloid}).  
We have used the embedding (\ref{embed2a})--(\ref{embed2e}) to 
convert the colliding bubble nucleation point 
$(T_{\rm c},R_{\rm c})$ into embedding coordinates.  The future
lightcone of our bubble is simply $X_4=\Hp^{-1}$.  It is 
straightforward to solve this system of equations for the angle
of their intersection, which gives
\beq
\cos(\theta) = \frac{\sin(T_{\rm c})\tan(T)+\cos(T_{\rm c})
-\cos(R_{\rm c})}{\sin(R_{\rm c})\,\tan(T)} \,,
\label{intersection}
\eeq
as a function of the closed dS chart time $T$ along the intersection.

\section{Distribution of observable bubble collisions}
\label{sec:distribution}

To compute the spatial distribution of bubble collisions on an 
observer's sky, we integrate over the spacetime volume available 
for colliding bubble nucleations, as a function of the coordinates
on the unit two-sphere parametrizing the observer's sky.  The 
differential number of bubbles nucleating as a function of the 
closed dS coordinates can be approximated
\beq
dN = \Gamma dV = \frac{\Gamma}{\Hp^4}\frac{\sin^2(R_{\rm c})}
{\cos^4(T_{\rm c})}\, dT_{\rm c}\,dR_{\rm c}\,
d\Omega_{\rm c} \,,
\label{dist}
\eeq
where $\Gamma$ is the bubble nucleation rate per unit 
four-volume $V$.  This is only an approximation because it treats
the four-volume in the future lightcone of any bubble nucleation
event the same as the four-volume of the parent vacuum, when in 
fact the geometry and the bubble nucleation rate will in general
be different there.  We return to this issue in Section 
\ref{sec:V1}.

The available volume for colliding bubble nucleations naturally 
divides into three sectors, labeled $V_1$, $V_2$, and $V_3$ in 
Figure \ref{fig:C}.  As described in Section \ref{sec:collision}, 
in this work we consider only colliding bubble nucleations in 
region $V_2$ to be observable.  Thus, we are interested in the 
spatial distribution of bubble collisions coming from there.  The 
corresponding volume is bounded in part by our bubble wall, which is 
approximated as the future lightcone of the center of our bubble 
nucleation event.  This gives the constraint
\beq
T_{\rm c} < R_{\rm c} \,.
\label{C1ab}
\eeq
The region $V_2$ is additionally bound by the requirement that 
(\ref{m&m}) has a real solution for the coordinate $\theta$ on the 
observer's sky; this gives the constraints
\beq
e^{\eta_\star-\xi_\star}
< (\Hd/\Hp)\tan\!\big[\textstyle\frac{1}{2}(T_{\rm c} + R_{\rm c})\big] 
< e^{\eta_\star+\xi_\star} \,.
\label{C23a}
\eeq    
Finally, $V_2$ is bounded by the initial hypersurface below which we 
presume there are no bubble nucleations.  According to our assumptions, 
this is given by either (\ref{h1b}) or (\ref{h2b}).

\begin{figure*}[t!]
\begin{center}
\includegraphics[width=0.33\textwidth]{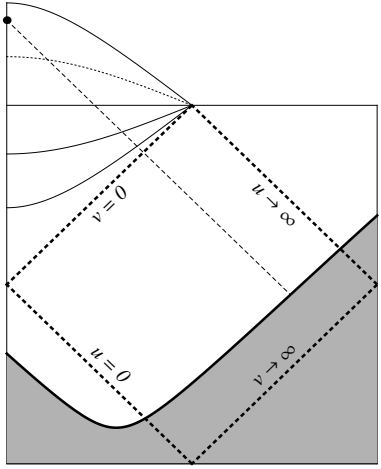} 
\caption{\label{fig:coords} Conformal diagram indicating surfaces of 
constant $u$ and $v$ (thick dotted lines).  The other features
are included to orient the diagram with respect to previous figures.}
\end{center}
\end{figure*}

These constraints suggest a new set of coordinates, 
$(T_{\rm c},\,R_{\rm c},\,\theta_{\rm c})\to (u,\,v,\,\zeta)$, where
\beq
u \equiv \tan\big[\textstyle\frac{1}{2}(T_{\rm c}+R_{\rm c})\big] \,,
\quad
v \equiv \tan\big[\textstyle\frac{1}{2}(T_{\rm c}-R_{\rm c})\big] \,,
\quad {\rm and} \quad
\zeta \equiv \cos(\theta_{\rm c}) \,.
\label{uvdef}
\eeq
Hypersurfaces of constant $u$ or constant $v$ are lightsheets, with $u=0$ 
corresponding to the past lightcone of our bubble nucleation event, 
and $v=0$ corresponding to its future lightcone, as indicated in 
Figure \ref{fig:coords}.  We take $\zeta$ to run from minus 
one to one, so that the (suppressed) coordinate $\phi$ runs from zero 
to $\pi$.  The region $V_2$ corresponds to intervals 
\bea
(\Hp/\Hd)\,e^{-\xi_\star} \,<\, &u& \,<\, 
(\Hp/\Hd)\,e^{\,\xi_\star} \label{urange}\\
\vmin(u) \,<\, &v& \,<\, 0 \,, \label{vrange}
\eea
where we have used $\eta_\star\ll\xi_\star$.  In the case of the 
initial hypersurface (\ref{h1b}), $\vmin$ is given by
\beq
\vmin = -\frac{\og(1+\ob\zeta)u+\beta}
{\beta u+\og(1-\ob\zeta)} \,,
\label{h1c}
\eeq
while in the case of the initial hypersurface (\ref{h2b}), it is
given by
\beq
\vmin = -\frac{\gamma\og\beta(1+\ob\zeta)u+\gamma-1}
{(1+\gamma)u+\gamma\og\beta(1-\ob\zeta)} \,.
\label{h2c}
\eeq
The parameters $\gamma$, $\beta$, $\og$, and $\ob$ are determined
by the proper time between the initial hypersurface and our bubble,
and the FRW radial coordinate $\xi_0$ of the observer in our bubble;
see Section \ref{sec:hypersurface}.   In terms of these coordinates, 
the differential volume element is given by
\bea
dV \!&=&\! \frac{2}{\Hp^4}\frac{(u-v)^2}{(1-uv)^4}\, 
du\,dv\,d\zeta\,d\phi_{\rm c} \,.
\label{dV2}
\eea

\subsection{Assumptions about model parameters}
\label{ssec:modelparameters}

Integrating the differential volume element (\ref{dV2}) within the 
regions delineated above is rather complicated.  It is therefore 
convenient to introduce some simplifying approximations.  To begin,
note that the current observational bound on the spatial curvature, 
$\Omega_{\rm curv}^0\leq 6.3\times 10^{-3}$ (WMAP+BAO+SN 95\% 
confidence level bound for $w=-1$ prior \cite{Komatsu:2010fb}), 
implies that the comoving distance to the surface of last 
scattering satisfies $\xi_\star\leq 0.27$.  We therefore expand in
$\xi_\star\ll 1$.  Meanwhile, the current observational bound on 
the amplitude of tensor perturbations on the surface of last 
scattering implies $G\Hd^2\lesssim 10^{-10}$ \cite{Komatsu:2010fb}.  
While it is not necessary that the parent vacuum have near 
Planck-scale energy density, this seems like a plausible assumption 
for typical states in the landscape, and appears to be required 
to permit the rapid decay rates we consider in this paper.  Therefore 
we also assume $\Hd/\Hp\ll 1$.    

To further simplify results, it is often assumed that the original
FRW radial coordinate of the unboosted observer, $\xi_0$, tends to 
infinity.  This is justified as follows.  The FRW symmetries of the 
bubble would seem to make any unit three-volume on a given 
constant--FRW time hypersurface as likely as any other to contain 
an observer.  At the same time, the hyperbolic geometry of the 
constant-FRW time hypersurface implies that the three-volume is 
dominated by regions at large $\xi$.  One would therefore expect a 
randomly-selected observer to reside at large $\xi$, with 
$\xi_0\to\infty$ in the limit of considering the entire bubble 
geometry.  However, these assumptions are problematic on multiple 
levels.  The result $\xi_0\to\infty$ itself points to an unregulated 
divergence, stemming from the diverging volume in the bubble.  
Regulating this divergence requires introducing a measure over the 
diverging volume of eternal inflation.  As it stands, it is unclear 
what the ``correct'' spacetime measure is, and different seemingly 
natural choices give very different cosmological predictions.  In 
the appendix we provide a detailed, if partial, review of the 
measure problem of eternal inflation.

At present, there is no viable measure that assigns equal 
likelihood (on average) for an observer to arise in any unit 
three-volume on a constant-FRW time hypersurface in the bubble, 
and is at the same time sufficiently precisely defined so as to 
predict, for example, the spatial distribution of bubble collisions
on an observer's sky.  Measures that have been understood to 
provide such a prediction have been found to suffer from Boltzmann
brain domination and runaway inflation (see the appendix).  
Moreover, it must be acknowledged that while the FRW symmetries 
that motivate such an approach hold to good approximation in some 
regions within a given bubble, they do not constitute a set of 
global symmetries on which a measure can be constructed, because 
of the effects of bubble collisions.   

The various measure proposals described in the appendix all point
toward the same qualitative prediction:  observers like us typically 
reside at $\xi_0\sim {\cal O}(1)$, with the distribution falling off 
exponentially with $\xi_0$ for $\xi_0$ greater than one.  Therefore, 
we take $\og=\cosh(\xi_0)\sim {\cal O}(1)$ and make the corresponding 
assumption for $\ob=\tanh(\xi_0)$.  We also take 
$\gamma=\cosh(\Hp\tau_{\rm n}/2)\gtrsim {\cal O}(1)$ and make the 
corresponding assumption for $\beta=\tanh(\Hp\tau_{\rm n}/2)$.  
Here $\tau_{\rm n}$ is the proper time between the nucleation of 
our bubble and the initial hypersurface (in the comoving frame of 
the parent vacuum).  Note that although we are interested in rapid 
vacuum decay rates, decays rates that are minuscule next to the 
Hubble rate of the parent vacuum would preclude the very notion of 
a metastable parent vacuum.  (A more precise statement of the 
assumption used below is $(\Hp/\Hd)\tanh(\Hp\tau_{\rm n}/2)\gg 1$, 
which might be seen to offer greater leeway toward considering 
rapid parent-vacuum decay rates.)  As a point of reference, note 
that the transition to eternal false-vacuum inflation occurs when
the total decay rate satisfies \cite{Guth:1982pn}    
\beq
\frac{\Gamma}{\Hp^4} \approx \frac{9n_c}{4\pi} \,,
\eeq
where $n_c$ is determined numerically to be $n_c\approx 0.34$.  
This corresponds to the rough expectation  
$\Hp\tau_{\rm n}\gtrsim (\Gamma/\Hp^4)^{-1} \approx 4.1$
at the transition to eternal inflation. 

Finally, consider the cosmology within our bubble.  This begins 
with a period of curvature domination, with the subsequent 
period of inflation beginning when $-\eta\approx 1$; recall the 
scale-factor solution (\ref{sf}).  Empirically, the number of 
$e$-folds of inflation is large, $\Ne\gtrsim 60$, meaning that by 
the end of inflation the FRW conformal time has become very small
in magnitude, $-\eta\approx 2e^{-\Ne} \lll 1$.  It can be shown that 
during its subsequent evolution, the conformal time $\eta$ becomes 
positive (since radiation domination lasts for a proper time interval 
$\Delta\tau\gg\Hd^{-1}$), but the total change in $\eta$ is 
dominated by its growth between recombination and the present 
(since the present FRW proper time is much greater than it was at
recombination), during which $\eta$ changes by an amount
\beq
\Delta\eta\equiv \xi_\star\sim\sqrt{\Omega_{\rm curv}^0} \,.
\label{rstar}
\eeq
Thus, $\eta_\star\ll\xi_\star$ and the present conformal time can 
be approximated
\beq
\eta_0=\eta_\star+\xi_\star\simeq\xi_\star \,.
\eeq
Because we work in terms of conformal time, and because we are 
only interested in the potential observability of bubble collisions
due to the causal structure of the multiverse, this is the only 
cosmology in our bubble on which we rely.

\subsection{Analysis of the distribution function}

It is straightforward to integrate (\ref{dV2}) with respect to 
$v$, and doing so we find
\beq
dN = \frac{2}{3} \frac{\Gamma}{\Hp^4}\,
\frac{\big[(u^4+u^2+1)\vmin^2-3(u^3+u)\vmin
+3u^2\big] \vmin}
{(u\vmin-1)^3}\,\,du\,d\zeta\,d\phi_{\rm c} \,. 
\label{dN1}
\eeq
Notice that the result is greatly simplified if we assume 
$u\gg 1$ and $|u\vmin|\gg 1$, in accordance with the assumptions
about model parameters outlined in Section \ref{ssec:modelparameters}
above (we discuss this approximation in greater detail below).  Then 
we find
\beq
dN \simeq \frac{2}{3} \frac{\Gamma}{\Hp^4} \,u\,du\,d\zeta\,
d\phi_{\rm c} \,, 
\label{dN2}
\eeq
independent of $\vmin$.  The independence of $\vmin$ indicates that
the spectrum of observable bubble collisions is dominated by 
collisions with bubbles that nucleate near $v=0$, i.e.~near the 
bubble wall of our bubble.  Likewise, the differential number of 
bubbles is isotropic with respect to the coordinates on the two-sphere 
of the observer's sky.  

The integral over $u$ is now trivial.  Before we report the result,
note that there is a simple correspondence between the coordinate 
$u$ and the angular scale of the region affected by the bubble 
collision on the observer's sky.  Defining the angular scale $\psi$ 
to be the largest value of $\theta-\theta_{\rm c}$ affected by 
the bubble collision, from (\ref{m&m}) we see
\beq
u = (\Hp/\Hd)\exp\!\big[\xi_\star\cos(\psi)+\eta_0-\xi_\star\big] \,.
\eeq
Thus, $dN\propto d\cos(\psi)$, i.e.~we obtain the standard result 
that the distribution of angular scales $\psi$ is uniform in 
$\cos(\psi)$.  Returning to (\ref{dN2}) and performing the 
integrations, we obtain
\beq
N \simeq \frac{16\pi}{3} \frac{\Gamma}{\Hp^4} \frac{\Hp^2}{\Hd^2}\, 
\xi_\star \,,
\label{Nresult}
\eeq
where we have expanded in $\xi_\star\ll 1$.  This confirms the 
well-known result, giving hope for observable bubble collisions if 
$\Gamma/\Hp^4$ and $\xi_\star$ are not too small.   

The comoving distance to the surface of last scattering 
falls off exponentially with the number of $e$-folds of
inflation in our bubble, $\xi_\star\propto e^{-\Ne}$.  In light 
of (\ref{Nresult}), one might worry that an overwhelming 
probability for $\Ne$ to be significantly larger than the current 
observational constraint makes the expected number of (observable) 
bubble collisions far below one.  While this is a valid concern, 
there are reasons to be hopeful.  The expected number of $e$-folds 
of inflation depends on the landscape distribution of $\Ne$ and on 
the measure regulating the diverging spacetime volume of eternal 
inflation.  In the first case, there is cause to 
expect that a large number of $e$-folds might be difficult to 
achieve in the string landscape, with the admittedly crude 
estimation of \cite{Freivogel:2005vv} indicating a distribution 
that falls off like $\Ne^{-4}$.  Meanwhile, measure proposals that 
weigh by the inflationary expansion factor $e^{3\Ne}$ in a bubble 
suffer from runaway inflation in a large landscape, giving cause to 
exclude them (see the appendix).  Indeed, none of the measure 
proposals described in Section \ref{ssec:measureproposals} give a 
preference for one bubble over another based on it having a larger 
number of $e$-folds of inflation.  Indeed in several cases the 
distribution of the curvature parameter has been predicted, 
conditioning on present observational constraints, and in each case 
one finds a roughly one in ten chance that 
$\Omega^0_{\rm curv}>10^{-5}$ (corresponding to $\xi_\star > 0.01$) 
\cite{Bousso:2009gx,DeSimone:2009dq,Salem:2011mj}.  

It is informative to reconsider the inequalities that lead to the
result (\ref{dN2}).  After restricting attention to ``observable'' 
bubble collisions, which limits $u$ to lie in the interval 
(\ref{urange}), the inequality $u\gg 1$ follows directly from our 
assumption $\Hd/\Hp\ll 1$.  For the moment take this as given, and 
consider the inequality $|u\vmin|\gg 1$.  In the case of the 
``tunneling from nothing'' initial hypersurface of (\ref{h1b}), 
$\vmin$ is given by (\ref{h1c}).  Then the inequality 
$|u\vmin|\gg 1$ can only fail if $\ob$ is very near to one.  This 
implies that $\og$ must be large, and the inequality fails either 
when $\og(1+\ob\zeta)u\lesssim 1$ or when 
$(1+\ob\zeta)u^2\lesssim 1$, depending on whether $u>\og$ or not
(recall that $\beta<1$).  
The first possibility actually does not admit a solution for 
$\zeta\geq -1$, while the second possibility holds only when 
$\zeta$ falls within a narrow window 
$\Delta\zeta \lesssim (\Hd/\Hp)^2$ near minus one (if also 
$\xi_0\gtrsim\ln(\Hp/\Hd)$).  The number of bubble collisions 
coming from this angular region in the observer's sky would be 
$\Delta N \lesssim (\Gamma/\Hp^4)\,\xi_\star$.  Even under our 
assumption of rapid decay rates, it is hard to accept that this 
number could be large enough to detect a significant deviation from 
the otherwise isotropic distribution of bubble collisions.  

A similar analysis applies to what we have called the ``bubble
wall'' initial hypersurface of (\ref{h2b}), for which $\vmin$ is
given by (\ref{h2c}).  In particular, assuming as before $u\gg 1$, 
the inequality $|u\vmin|\gg 1$ fails only when $\ob$ is very near
to one, and even in that case only when $\zeta$ lies within a 
narrow window near minus one.  The specifics differ only in the
placement of factors of $\beta$ and $\gamma$, and assuming the 
former isn't too small (i.e.~assuming $\beta u\gg 1$), then we 
reach the same qualitative conclusion:  the anisotropy in the 
spectrum of bubble collisions is limited to an angular region
$\Delta\zeta\lesssim (\Hd/\Hp)^2$ (if also 
$\xi_0\gtrsim\ln(\Hp/\Hd)$), from which a negligible number of 
bubble collisions are available to experimentally resolve the 
anisotropy.  

These arguments fail when $u$ is not large, i.e. when $\Hd\sim\Hp$.  
In this case the integration over $u$ covers only a small interval 
$\Delta u \simeq 2\,(\Hp/\Hd)\,\xi_\star$, which allows us to 
approximate the integral by keeping the argument fixed, 
evaluated at the mean value $\u\equiv (\Hp/\Hd)$.  This gives
\beq
dN = \frac{4}{3} \frac{\Gamma}{\Hp^4}
\frac{\big[(\u^4+\u^2+1)\vmin^2-3(\u^3+\u)\vmin
+3\u^2\big]\u\vmin\,\xi_\star}
{(\u\vmin-1)^3}\,\,d\zeta\,d\phi_{\rm c} \,. 
\label{dN3}
\eeq
By inspecting (\ref{h1c}) and (\ref{h2c}), we see that the 
magnitude of $\vmin$ is never much greater than one (given the
present hypothesis that $\u$ is not much greater than one), 
and at the same time $\u$ is always positive and $\vmin$ is 
always negative, so (\ref{dN3}) does not feature any poles.  
Thus we come back to the conclusion from above:  although $dN$
now features anisotropy over a broad angular region, the total
number of bubble collisions coming from this region is small,
and therefore the anisotropy is unobservable.  We note that 
this is the same effect found by FKNS in their study of the
GGV ``persistence of memory,'' in the case where the initial
hypersurface is pushed far into the past of our bubble.  

The above analysis considers only two orientations of the 
initial hypersurface.  However it seems reasonable to suppose
that insofar as the semi-classical multiverse can be described 
in terms of evolution from an initial hypersurface below which 
there are no bubble collisions, that such a hypersurface will 
thread somewhere between the two orientations that we consider, 
in which case we expect the same results.  Thus we conclude that 
if the effects of a significant number of bubble collisions are 
observable, the distribution of these collisions on an observer's 
sky will appear isotropic, with the angular scale $\psi$ of 
collision regions distributed uniformly with respect to 
$\cos(\psi)$, regardless of our proximity to any surface 
specifying initial conditions.

\section{``Unobservable'' bubble collisions}
\label{sec:V1}

In studying the distribution of bubble collisions in Section 
\ref{sec:distribution}, we ignored colliding bubbles coming from 
the region $V_1$ in Figure \ref{fig:C}, because as explained 
in Section \ref{sec:collision} the causal futures of these collisions 
cover the entire surface of last scattering.  However, although such 
bubble collisions are likely to be difficult to observe directly, 
they can have important consequences.  If, for example, the number
of these bubble collisions is very large, then it also seems the 
probability is large for there to be intrusive domain walls such as 
in the right panel of Figure \ref{fig:C2}.  As remarked in Section 
\ref{sec:collision}, this would not necessarily disagree with our 
observation of a (nearly) flat FRW Hubble volume, but it would seem 
to imply that selection effects play an important role in placing 
observers like us at special locations in their bubbles.  

To begin, we compute the expected number of bubble collisions coming
from region $V_1$.  This region corresponds to the coordinates $u$ 
and $v$ of (\ref{dV2}) lying in the intervals
\bea
0 \,<\, &u& \,<\, (\Hp/\Hd)\,e^{-\xi_\star} \label{urange2}\\
\vmin(u) \,<\, &v& \,<\, 0 \,, \label{vrange2}
\eea
where $\vmin$ is given by (\ref{h1c}) or (\ref{h2c}), depending
on the choice of initial hypersurface.  The range of $v$ has the 
same form as in region $V_2$, and so we again obtain (\ref{dN1}).  
Likewise, in the subset of the parameter space in which $u\gg 1$  
and $|u\vmin|\gg 1$, the expression simplifies to (\ref{dN2}).  
These inequalities are invalidated at the lower limit of the 
$u$ integration, where $u\to 0$, but let us ignore that for the 
moment.  The expression (\ref{dN2}) is then easily integrated
to give
\beq
N \simeq \frac{4\pi}{3} \frac{\Gamma}{\Hp^4} \frac{\Hp^2}{\Hd^2}\,,
\label{NV1}
\eeq  
where we have expanded in $\xi_\star$.  As before, the inequality 
$|u\vmin|\gg 1$ can fail even when $u$ is large, if $\zeta$ falls 
into a narrow window $\Delta\zeta\sim 1/u^2$, but by the same 
arguments given above this region of parameter space does not 
contribute significantly toward $N$.  This leaves the subset of 
the integration space where the inequality $|u\vmin|\gg 1$ fails 
because $u$ is small.

If we take $\xi_0\sim{\cal O}(1)$, then $\og=\tanh(\xi_0)$ is not 
extremely close to one, and consequently for either initial 
hypersurface $|\vmin|$ is well-behaved (that is, of order $u$ or 
of order unity) as $u$ is taken to zero.  Meanwhile, since $u\vmin$ 
is always negative, the magnitude of the denominator of (\ref{dN1}) 
is always greater than one.  Therefore the distribution (\ref{dN1}) 
is dominated by large values of $u$, as is the integration over $u$, 
validating the result (\ref{NV1}).  

If instead we consider large values of $\xi_0$, then $\ob$ can 
become very close to one and $|\vmin|$ can grow large in 
the limits of integration where $u\to 0$ and $\zeta\to\pm 1$.  Given 
our setup, it is complicated to carefully analyze this limit.  
Nevertheless, in the limit of small $u$ and large $|\vmin|$, the 
distribution (\ref{dN1}) becomes proportional to 
$(u-\vmin^{-1})^{-3}_{\phantom{i}}$, which can be integrated and
evaluated in the above limits.  We find that when 
$\xi_0\gg\ln(\Hp/\Hd)$, $N$ receives a contribution that grows 
linearly with increasing $\xi_0$, 
$\Delta N \approx (2\pi/3)\,(\Gamma/\Hp^4)\,\xi_0$.

Section \ref{ssec:modelparameters} argues that the distribution of
observers falls off exponentially with $\xi_0$, and so a typical
observer should expect the result (\ref{NV1}).  If one resists 
those arguments and instead takes observers to arise at arbitrarily
large $\xi_0$, then at first glance it seems that a typical observer 
would have an arbitrarily large number of bubble collisions in his 
past lightcone.  On the other hand, in a sufficiently large landscape 
there is a finite probability for any random bubble collision to 
create a domain wall that moves toward the center of the bubble, 
disrupting the open FRW symmetry in its wake, and in particular 
supplanting some of the would-be large-$\xi_0$ volume in the 
observer's bubble with the vacuum type of the colliding bubble 
(as in the right panel of Figure \ref{fig:C2}).  The probability to
have encountered such a disruptive bubble collision in the past 
lightcone grows with $\xi_0$, yet so does the would-be available 
three-volume on a fixed (open) FRW time hypersurface in the bubble.  
It can be shown that, after accounting for the volume subtracted by 
disruptive bubble collisions, the three-volume still diverges at 
large $\xi_0$ (if inflation is eternal in the parent vacuum); only it 
does so along special angular directions that by chance have not 
experienced any significantly disruptive bubble collisions in their 
past lightcones (see for example \cite{Dahlen:2008rd}).  

Returning to the assumption $\xi_0\sim {\cal O}(1)$, the number
of bubble collisions coming from the region $V_1$ is given by 
(\ref{NV1}), which is ``larger'' than the number of observable 
bubble collisions (\ref{Nresult}) by a factor of $1/4\xi_\star$.  
Given the constraint $\xi_\star\leq 0.27$, this factor is not 
necessarily very large, but one can imagine parameters for which 
the number of observable bubble collisions is significant and at 
the same time the number of unobservable bubble collisions is 
much larger.  If one or more of these latter collisions occurs 
sufficiently early and creates a domain wall that moves toward 
the center of our bubble, then it can significantly deform the 
rotational symmetry of our bubble wall.  For example, the 
collision illustrated in the right panel of Figure \ref{fig:C2} 
can be seen to create a large ``dimple'' in the would-be 
two-sphere corresponding to our bubble wall (at some fixed time
after the collision).  If a significant number of bubble 
collisions come from bubbles that nucleate within this first 
bubble (i.e.~within the lightly-shaded region in Figure 
\ref{fig:C2}), then their distribution across the observer's sky 
will not be isotropic, due to the dimple deformation.  
Furthermore, it is frequently argued that bubble collisions do 
not generate significant levels of gravitational radiation, due 
to a generalization of Birkhoff's theorem applied to the 
SO(2,1)-symmetric geometry in the wake of the collision 
\cite{Kosowsky:1991ua}.  Yet, for the ``bubble-in-bubble'' 
collision sequences described above, there is no symmetry 
argument precluding the generation of significant gravity waves.

Note that domain walls move toward the center of our bubble only
when the vacuum energy behind the domain wall is less than that
in our bubble \cite{Chang:2007eq}.  Since the instantons describing
the decay of low-energy vacua are likely to have large (Euclidian) 
actions, these colliding bubbles are unlikely to produce many 
colliding bubble nucleations within them.  Indeed, it might be 
seen that the parent vacuum does not decay to any dS vacua with 
lower energy than ours (the relevant energy scale in our bubble is 
the inflationary scale, not the scale of the present-day 
cosmological constant, however even the former could be unusually 
small in magnitude among states in the landscape).  On the other 
hand, perhaps even the domain walls of large vacuum-energy 
colliding bubbles sufficiently break the symmetries of our bubble 
wall for subsequent collisions to generate significant gravity 
waves.    

It is difficult to gauge the likelihood of these possibilities, 
but we suggest the following calculation as an intuitive guide.  
As we have remarked, at any given time the bubble wall of our 
bubble is a two-sphere, and the domain walls of previous bubble 
collisions can be seen as ``dimples'' on this two-sphere.  We 
estimate the solid angle subtended by these dimples, at the last 
moment that the bubble wall of our bubble is within our past 
lightcone.  (At our level of approximation, which expands in 
$\xi_\star\ll 1$, this is equivalent to estimating the solid 
angle subtended by these dimples at the time right before bubble
collisions become ``observable.'')  The hope is that this captures 
an essence of the likelihood that a given bubble collision occurs 
in the wake of another one (since such collisions would correspond 
to ``dimples in dimples'' on the bubble wall).  Note that this 
cannot be the precise quantity that we wish it to be, for one 
because the solid angle subtended by a given dimple is independent 
of the vacuum energy and the decay rate within the colliding 
bubble that produced it, yet these quantities are important for 
determining the likelihood of bubble-in-bubble collisions.

The angular size of the dimple coming from a single bubble 
collision is given by (\ref{intersection}).  We are interested 
in evaluating this expression at the intersection of the bubble 
wall with our past lightcone, which corresponds to the closed dS time 
$T=\arctan\!\big[(\Hp/\Hd)e^{\xi_\star}\big]\simeq\arctan(\u)$,
where $\u\equiv \Hp/\Hd$ as before.  In terms of the coordinates 
$u$ and $v$ of (\ref{uvdef}) and the solid angle 
$\Omega_{\rm coll} = 2\pi\big[1 - \cos(\theta)\big]$, we write
\beq
\Omega_{\rm coll} = \frac{4\pi(u-\u)\,v}{(u-v)\,\u} \,.
\eeq 
We estimate the fraction of the total solid angle subtended by 
bubble collisions as
\beq
f = \frac{1}{4\pi} \int dV \,\Omega_{\rm coll}\,\Gamma \,,
\eeq
where $dV$ is given by (\ref{dV2}), and the range of integration
covers $0<u<\u$ and $\vmin<v<0$.  Note that this is an overestimate, 
because it double-counts the solid angles in bubble-in-bubble 
collisions.  The factor of $u-\u$ interferes with the approximation 
methods we have used so far.  Therefore, we instead note that all 
of our calculations for $\xi_0\sim {\cal O}(1)$ have been consistent 
with the results for $\xi_0=0$, to leading order in $\u$.  Therefore 
we approximate the above integral using $\vmin$ evaluated at 
$\xi_0=0$, i.e.~we set $\ob=0$ and $\og=1$.  Expanding the result to 
leading order in $\Hd/\Hp\ll 1$, we find for both initial hypersurfaces
\beq
f \simeq \frac{2\pi}{3}\frac{\Gamma}{\Hp^4}\ln(\Hp/\Hd) \,.
\label{fraction}
\eeq
(For the ``tunneling from nothing'' initial hypersurface, the 
correction to the logarithm is a $\tau_{\rm n}$-dependent term that 
never exceeds magnitude 1/2.  For the ``bubble wall'' initial 
hypersurface, the corresponding correction is 
$\ln\!\big[\!\tanh(\Hp\tau_{\rm n}/4)\big]-{\rm sech}^2(\Hp\tau_{\rm n}/4)$,
which can become important when $\Hp\tau_{\rm n}\lesssim 1$.  
This also suggests that the result for this case could be 
significantly changed by reintroducing $\xi_0\ne 0$.  Of course, the 
exact expressions in each case are always positive.  The ``fraction'' 
(\ref{fraction}) can be greater than one because of double counting.)

If the decay rate of the parent vacuum is hardly suppressed in 
Hubble units, then for large $\Hp/\Hd$ the fraction of the bubble
wall available for bubble-in-bubble collisions can be significant,
however it seems more plausible that this fraction is small.  
Furthermore, note that since the total number of colliding bubbles 
is given by (\ref{NV1}), the solid angle subtended by any single 
collision is on average 
$4\pi f/N\sim 2\pi\,(\Hd/\Hp)^2\ln(\Hp/\Hd)\ll 1$.  
This is indication that most of the bubble collisions occur at
late times, subtending a minuscule fraction of the total bubble 
wall in our past lightcone (do not confuse this with the fraction 
of the cosmic microwave sky affected by the bubble collision), and 
that an appreciable value of $f$ would be due primarily to a very 
large number of bubble collisions.

\section{Conclusions}
\label{sec:conclusions}

Spacetime on very large scales could feature false-vacuum eternal 
inflation, in which case our local Hubble volume is part of an 
infinite, open FRW universe contained within a CDL bubble.  Our 
bubble will generically collide with other bubbles, and under 
appropriate circumstances the effects of a bubble collision could 
leave observable imprints on, for example, the cosmic microwave 
background, if the collision occurs within our past lightcone.     

The potential to observe such bubble collisions, including their
spatial distribution across the CMB sky, the distribution of 
angular scales of collision-effected regions on the sky, and 
various other observational signatures, has been studied 
extensively in the literature (see the introduction).  To predict 
the spatial distribution of bubble collisions one must choose a 
measure to regulate the diverging volume of the eternally-inflating 
multiverse, including choosing an initial hypersurface below which 
there are no bubble nucleations.  This initial hypersurface defines 
a comoving frame for the multiverse, and Garriga, Guth, and Vilenkin 
found that this frame can in principle affect the spatial distribution 
of bubble collisions across an observer's sky, even in the limit where 
the initial hypersurface is pushed arbitrarily far into the past.  On 
the other hand, Freivogel, Kleban, Nicolis, and Sigurdson (FKNS) 
found that this information about the initial hypersurface is 
effectively screened when the inflationary Hubble rate within our 
bubble is much smaller the expansion rate outside, $\Hd/\Hp\ll 1$.    

We study the effect of this initial hypersurface under more 
general considerations, in particular envisioning our bubble to be 
very near to it.  We find the conclusion of FKNS to be robust; so
long as $\Hd/\Hp\ll 1$, any effects associated with the placement
and orientation of the initial hypersurface are relegated to an 
angular region $\Delta\cos(\theta)\lesssim (\Hd/\Hp)^2$, over which 
the expected number of bubble collisions is less than the 
dimensionless decay rate of the parent vacuum.  We show this 
explicitly for two choices of initial hypersurface, however the 
analysis suggests a more generic result.  This is because the 
vast majority of colliding bubbles nucleate very near to our bubble, 
and at the latest times consistent with their being observable.  We
review of the measure problem of eternal inflation in the appendix.

\acknowledgments

The author thanks Adam Brown, Ben Freivogel, Daniel Harlow, 
Shamit Kachru, Radford Neal, Steve Shenker, and Vitaly Vanchurin 
for helpful discussions.  The author is also grateful for the 
support of the Stanford Institute for Theoretical Physics, and for 
the hospitality of the Perimeter Institute, where some of this 
work was completed.

\appendix

\section{The measure problem:  a partial review}
\label{sec:measure}

We here briefly review the measure problem of eternal inflation.  
The goal of this appendix is to provide context and justification 
for various claims asserted in the main text.  As such, we do not 
provide a thorough overview of the subject; rather we focus on 
issues that relate to the phenomenology of bubble collisions.  
Likewise, references are included to point toward proximate 
supporting literature, and not to provide a proper historical 
account of original work.  For other, more complete reviews see 
for example \cite{Winitzki:2006rn,Vilenkin:2006xv,Guth:2007ng,
Freivogel:2011eg}.

Eternal inflation generates diverging spacetime volume.  In a 
theory of multiple metastable states, this translates to a 
diverging number of bubble nucleations.  Environmental variables, 
such as the local vacuum energy density, or the distribution of 
high- and low-temperature perturbations to the local cosmic 
microwave background, take all possible values an endless number 
of times.  A proposal for how to make predictions is called a 
measure.  It is worth emphasis that a measure is a necessary 
prerequisite to any probabilistic prediction---the measure is just 
the probability space---and that the standard ``Born rule'' measure 
of quantum mechanics is by itself insufficient in the context of 
diverging spacetime volume \cite{Page:2009qe}.

\begin{figure}[t!]
\begin{center}
\includegraphics[width=0.50\textwidth]{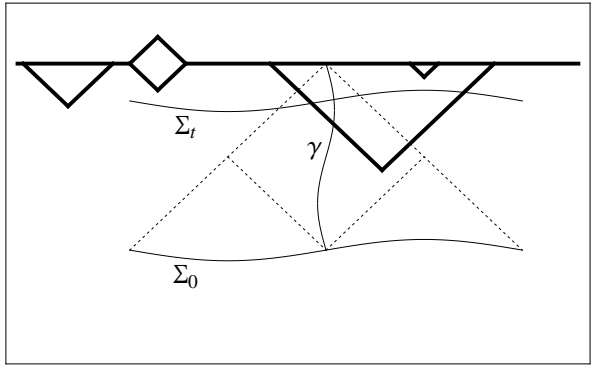}
\caption{\label{fig:multiverse} A cartoon conformal diagram
of a subset of a single, semi-classical realization of 
spacetime.  The thick, mostly-horizontal line is conformal 
infinity, and the thick future lightcones represent bubble 
nucleations.  The diagram displays four bubbles, but the 
spacetime is actually a fractal, with countless bubbles 
appearing toward future infinity.  The other features are 
explained in the main text.}
\end{center}
\end{figure}

Figure \ref{fig:multiverse} provides a cartoon conformal diagram 
of a subsection of a single, semi-classical realization of 
spacetime.  ``Semi-classical'' here refers to an understanding that 
the spacetime evolves classically, but with probabilistic quantum 
transitions (such as bubble nucleations) inserted randomly by hand.  
A spacelike hypersurface $\Sigma_0$ is indicated, along with a 
worldline $\gamma$.  The dotted lines correspond to the boundaries of 
the past lightcone and causal diamond of $\gamma$, treating 
$\Sigma_0$ as an initial hypersurface.  From the perspective of 
quantum theory, we consider this realization of spacetime as one 
among an ensemble, the elements of the ensemble corresponding to 
the various possibilities for the locations and types of bubble
nucleations (and other probabilistic quantum transitions), weighted 
by the branching ratios implied by the various tunneling (and 
other) transition rates.  We refer to these as ``future 
histories.''  

Figure \ref{fig:multiverse} displays a subset of the multiverse 
populated by CDL transitions to bubbles of positive vacuum energy 
(dS bubbles for short), negative vacuum energies (AdS bubbles for 
short), and precisely zero vacuum energy (Minkowski bubbles for 
short), all within an inflating false-vacuum background.  The dS 
bubbles feature eternal inflation in their interiors; meanwhile 
Minkowski bubbles evolve toward null future infinity and AdS 
bubbles end in a spacelike singularity \cite{Coleman:1980aw}.  
Note that any of these bubbles could feature slow-roll inflation
followed by reheating and big bang evolution.
In addition to the false-vacuum eternal inflation above, the 
multiverse could in general feature stochastic eternal inflation 
\cite{Vilenkin:1983xq,Linde:1986fd}, which occurs when scalar field
potentials are sufficiently shallow and field distributions 
sufficiently smooth that the random quantum fluctuations of fields 
as modes exit the Hubble radius during inflation compete with 
the classical evolution of such fields toward some local minimum.  
To simplify the discussion we set aside this possibility; it is 
straightforward to generalize.

With this view of the multiverse in mind, we describe two common
approaches toward constructing a measure.  One approach focuses on 
the (semi-classical) evolution of a finite, spacelike, ``initial'' 
hypersurface, for instance $\Sigma_0$ in Figure \ref{fig:multiverse}.  
If $\Sigma_0$ contains an eternal worldline, its future evolution 
contains diverging spacetime volume.  Nevertheless, given a 
foliation $t$ (with for example $t=0$ on $\Sigma_0$), one can 
consider the finite spacetime volumes between $\Sigma_0$ and the 
constant-$t$ hypersurfaces $\Sigma_t$.  Calculations proceed by 
assuming the spacetime obtained by restricting attention to the 
sample between $\Sigma_0$ and $\Sigma_t$ faithfully represents the 
whole, in the limit $t\to\infty$.  For example, to compute the 
relative probability of experimental outcomes $A$ and $B$, one 
studies a future history between $\Sigma_0$ and $\Sigma_t$, counting 
the number of times $N_A$ the experiment leads to $A$, and the 
number of times $N_B$ it leads to $B$.  The relative probability is 
then the ratio, in the limit $t\to\infty$,
\beq
\frac{P_A}{P_B} = \lim_{t\to\infty}\, \frac{N_A}{N_B} \,.
\eeq
One could in principle expand the program to include an ensemble of 
initial conditions on $\Sigma_0$, and/or the ensemble of semi-classical 
future histories of $\Sigma_0$.  However, in the cases of interest it is 
argued that the statistics of such ensembles are faithfully captured by 
the above limiting procedure, applied to a single (infinite) 
semi-classical realization of the spacetime.  This class of measure 
proposals is referred to as ``global'' measures.  Examples of global 
measures include the proper-time cutoff measure \cite{Vilenkin:1994ua}, 
which takes $t$ to be the proper time along a congruence of geodesics 
initially orthogonal to $\Sigma_0$, and the scale-factor cutoff measure 
\cite{DeSimone:2008bq,DeSimone:2008if,Bousso:2008hz}, which take $t$ to 
reflect the integrated expansion along such a congruence (see below).

The second approach centers the analysis on the future histories 
surrounding a single worldline, for instance $\gamma$ in Figure
\ref{fig:multiverse}.  In particular, the measure provides a rule
for assigning a finite spacetime volume to a finite worldline in
a given semi-classical realization of spacetime, and then expands
the program to include all future histories surrounding the 
worldline, each future history weighted according to the branching 
ratio implied by any quantum transitions occurring in the spacetime.  
(The proposals tend to remain uncommitted to any rule for weighting
across different initial conditions, on which the predictions of
these measures generically depend \cite{Vanchurin:2006xp}, though 
such a rule could be selected arbitrarily from existing proposals 
\cite{Hartle:1983ai,Vilenkin:1984wp}).  For example, in the causal 
diamond measure the rule is to count only those events that occur 
within the causal diamond of the given worldline.  In the practical 
terms used above, the relative probability of experimental outcomes 
$A$ and $B$ is computed by counting the number of experiments 
leading to $A$ and to $B$ within the causal diamond of the 
worldline in a given semi-classical realization of spacetime, 
multiplying these numbers by the branching ratio of that realization 
of spacetime, summing these numbers over all semi-classical 
realizations, and performing a weighted sum over initial conditions, 
taking the ratio in the end.  Such a program would seemingly fail 
for an eternal worldline; however such worldlines form a subset of 
measure zero.  (Put another way, the branching ratios of 
semi-classical histories in which the worldline avoids terminating 
on an AdS singularity decrease exponentially with increasing 
worldline length.  At the formal level, any stable Minkowski vacua 
must be ignored.)  Such measures are referred to as ``local'' 
measures.  Examples of local measures include the causal patch 
measure \cite{Bousso:2006ev} and fat geodesic measure 
\cite{Bousso:2008hz} (see below).

Note that both of these approaches to the measure problem establish 
a preferred frame in each semi-classical realization of spacetime.  
In the global measures, this frame is defined by the initial 
hypersurface $\Sigma_0$, or, alternatively, the orthogonal 
congruence of geodesics that emanate from it.  In the local measures,
this frame is defined by the specification of initial conditions for 
the semi-classical histories surrounding the worldline $\gamma$ 
(including the orientation of the initial tangent of $\gamma$ with 
respect to those initial conditions).

In what follows, we frequently refer to ``observers like us'' and to
bubbles/vacua ``like ours.''  Unless otherwise noted, these are 
intended as precise conditioning statements, focusing in on observers 
who share our beliefs about the world in which we live, and what those 
beliefs imply about that world given the theoretical context outlined 
above.  Such precise conditioning is appropriate for making predictions 
(as opposed to postdictions), and also serves to simplify the 
discussion.  More general considerations are discussed in 
\cite{Garriga:2007wz}.  As in the main text, we everywhere work in 
3+1 spacetime dimensions.

\subsection{Phenomenological pathologies}
\label{sec:pathologies}

\subsubsection{Youngness paradox}

Before describing a few measures in greater detail, we take a moment
to discuss some of the criteria used to favor these measures over 
others.  To begin, we study the youngness paradox of the proper-time
cutoff measure.  This refers to the prediction that it is 
overwhelmingly more likely that we would have arisen at some earlier 
FRW time than the present (and for example measured some higher CMB 
temperature than 2.73 K) \cite{Tegmark:2004qd,Guth:2007ng,Bousso:2007nd}.

For simplicity, imagine that there is only one vacuum like ours 
in the landscape, and it can arise (via CDL bubble nucleation) in 
only one type of parent vacuum, with dS curvature radius $\Hp^{-1}$.  
An ``initial'' ($t=0$) hypersurface $\Sigma_0$ defines a proper-time 
foliation $t$, and if $\Sigma_0$ intersects 
an eternal worldline, then for sufficiently large $t$, a 
constant-$t$ hypersurface $\Sigma_t$ will intersect many bubbles 
containing our vacuum.  We focus on two subsets of these bubbles: 
``early'' bubbles, which nucleate a proper time $t_{\rm early}$ 
below $\Sigma_t$, and ``late'' bubbles, which nucleate a proper time 
$t_{\rm late}$ below $\Sigma_t$, with 
$\Delta t = t_{\rm early}-t_{\rm late}>0$.  
Then the volume available for late bubble nucleations is larger than 
that available for early bubble nucleations by a factor 
$\sim e^{3\Hp\Delta t}$, meaning the number of late bubbles is larger 
than the number early bubbles by that factor.  Although the early 
bubbles expand for more time before $\Sigma_t$ than the late bubbles 
(both in terms of their bubble wall trajectories and in terms of 
their scale-factor evolution), if $\Delta t\gg\Hp^{-1}$, and if the 
Hubble rate in our bubble is significantly smaller than $\Hp$, then 
both of these effects are negligible next to the factor 
$e^{3\Hp\Delta t}$ \cite{Bousso:2007nd}.    

To draw just one consequence of this, consider the physical 
three-volume at two temperatures, $T_{\rm low}$ and $T_{\rm high}$ 
($T_{\rm low}<T_{\rm high}$), in each subset of bubbles.  Since early 
bubbles nucleate with more time below $\Sigma_t$, regions below 
$\Sigma_t$ in their interiors can reach lower temperatures, and we 
set $T_{\rm low}$ so that there are regions in early bubbles that 
reach this temperature, but not so in late bubbles.  Then the 
statement of the previous paragraph is this:  although the 
three-volume below $\Sigma_t$ at temperature $T_{\rm low}$ in a given 
early bubble can be greater that at temperature $T_{\rm high}$ in 
a given late bubble, the ratio of these volumes is negligible next to 
the factor of $e^{3\Hp\Delta t}$ accounting for the greater number of 
late bubbles, if the time required to cool from $T_{\rm high}$ to 
$T_{\rm low}$ is much greater than $\Hp^{-1}$.  In bubbles like ours, 
regions with a CMB temperature of 2.75 K correspond to 150 million 
years before the present FRW time and, after accounting for the 
greater number of late bubbles, would occupy $\sim e^{10^{48}}$ times
more volume than regions at 2.73 K, for GUT-scale $\Hp$.  While it 
might be less probable per unit volume for observers otherwise like 
us to arise at that time, it is unclear how such an effect could 
possibly compensate this enormous volume factor.  Note that although 
our discussion refers to a specific hypersurface $\Sigma_t$, the 
results hold in the limit $\Sigma_{t\to\infty}$.  Thus we conclude 
that the proper-time cutoff measure is inconsistent with the world 
we observe.

\subsubsection{Runaway inflation}

At first thought, it might seem natural for the measure to 
weight spacetime regions in such way that regions in the wake
of slow-roll inflation receive additional weight due to the 
volume expansion factor $e^{3\Ne}$.  In a sufficiently large landscape,
however, this leads to the problem of runaway inflation, where
one expects cosmological parameters related to the inflationary
history to take values very different than the ones we 
observe \cite{Feldstein:2005bm,Garriga:2005ee,Graesser:2006ft}.  

Consider for example a landscape of vacua in which both the number 
of $e$-folds of slow-roll inflation $\Ne$ and the primordial density
contrast $Q$ depend parametrically on some variable $\sigma$, which 
takes an effectively continuous range of values.  For example, 
$\sigma$ could be a coupling specifying the self-interaction of the 
inflaton field.  If the measure weights regions according to their
volume expansion factor, then we can write the regulated 
distribution of $\sigma$
\beq
{\cal N}(\sigma) \sim \rho_{\rm obs}\big[Q(\sigma)\big]\, 
{\cal P}(\sigma)\, e^{3\Ne(\sigma)} \,. 
\label{Nfactorization}
\eeq
Here ${\cal N}(\sigma)$ denotes the number of observers who would 
measure a given value of $\sigma$, $\rho_{\rm obs}$ is the average 
density of observers, which for simplicity we assume to depend on
$\sigma$ only via its dependence on $Q$ (we comment on this below),
and ${\cal P}$ is the volume the measure assigns to vacua with a 
given value of $\sigma$, modulo the volume-expansion factor that has 
been factored out.  (A given measure will not always lend itself to 
an intuitive factorization of this form, but it is straightforward 
to translate the argument to more specific proposals.)

Observers reside predominantly in regions where $\sigma$ is 
near a peak in ${\cal N}$, i.e.~in regions where $\sigma$ is valued 
near a solution of $d{\cal N}/d\sigma=0$.  Writing this out explicitly,
\beq
\frac{dQ}{d\sigma}\frac{d\ln\rho_{\rm obs}}{dQ}
+\frac{d\ln{\cal{P}}}{d\sigma}+3\Ne\frac{d\ln\Ne}{d\sigma} =0 \,.
\label{diffeq}
\eeq  
There is no reason to expect the distribution ${\cal P}(\sigma)$ to
have precisely the exponential dependence on $\sigma$ required to
offset the last term in (\ref{diffeq}), at values of $\sigma$ near
an anthropic window.  This implies that the cancellation
of the last term comes almost entirely from the first term.  And
since $Q$ depends parametrically on $\sigma$, this means 
$\rho_{\rm obs}(Q)$ should depend exponentially on $Q$ (we 
emphasize this is a very strong exponential dependence, since 
$3\Ne\gtrsim {\cal O}(100)$ in our bubble).  Yet such a strong
exponential dependence is in conflict observation, given for 
instance that the value of $Q$ is such that galaxies like ours 
are rather typical, as opposed to exponentially unlikely.  (The
dependence of $\rho_{\rm obs}$ on $Q$ is studied more carefully
in \cite{Tegmark:1997in}.)   

Note that we encounter the same issue if the (3+1)-dimensional 
gravitational constant $G$ depends parametrically on some parameter
that scans effectively continuously across the landscape (for 
example, a modulus coupling determining the volume of compact 
dimensions).  This is because $\Ne$ and $Q$ generically depend 
parametrically on $G$ (regardless of the inflationary parameters), 
which is itself a cosmological parameter that in our universe 
evidently sits comfortably within its anthropic window 
\cite{Graesser:2006ft}.  

This argument is not airtight.  For instance in the example above,
$\rho_{\rm obs}$ could depend on $\sigma$ more than just via its 
dependence on $Q$.  Then there could be some ``hidden'' (sharp) 
anthropic dependence on $\sigma$, for instance in dynamics associated 
with baryo/leptogenesis \cite{Hall:2006ff}.  However this would have 
to be the case for all runaway parameters (all effectively 
continuously scanning parameters on which $\Ne$, $Q$, and $G$ depend), 
which seems very unlikely in an enormous landscape such as the 
string landscape.  It is also possible that for every potential 
runaway parameter $\sigma$, ${\cal P}(\sigma)$ has a sharp feature 
at a value of $\sigma$ that coincides with sitting comfortably 
within the anthropic range.  This too seems unlikely in an enormous landscape, but it is admittedly hard to anticipate whether allowing 
a large number of additional parameters to scan might allow for a 
large number of anthropic windows, some of which coincide with the 
extremal values of the runaway parameters.

\subsubsection{Boltzmann brains}

CDL bubble nucleations are presumably not the only quantum transitions
allowed in locally--dS space.  The possibility to nucleate an
observer (along with any supporting environment deemed necessary to 
properly condition the predictions of the observer) directly out of 
an otherwise empty, positive-energy vacuum poses a challenge to the 
standard view of cosmology:  if such observers, generically referred 
to as ``Boltzmann brains'' (BBs), are predicted to vastly outnumber 
the observers arising in the wake of hot big-bang evolution following 
reheating (``normal observers,'' or NOs), then our observations appear 
very atypical, calling into question the theoretical premise of the 
original prediction \cite{Dyson:2002pf,Albrecht:2004ke,Page:2006dt,
Bousso:2006xc,DeSimone:2008if}.  It is worthwhile to spell this out 
explicitly with a concrete example.

For simplicity imagine that within the eternally-inflating multiverse 
there is only one vacuum with the low-energy degrees of freedom 
necessary to create observers of any type, and it resembles ours.  
Then, roughly speaking, the measures described below predict that 
the balance of BBs and NOs depends on whether the decay rate 
$\Gamma_{\rm vac}$ of the anthropic vacuum is greater 
than or less than the BB formation rate $\Gamma_{\rm BB}$, with BBs 
dominating when $\Gamma_{\rm BB} > \Gamma_{\rm vac}$.  Suppose we are 
confident in the theoretical setup, and only uncertain as to the decay 
rate $\Gamma_{\rm vac}$.  Before considering the issue of BBs, we 
divide the space of theories into two classes, $T_{\rm NO}$ in which 
$\Gamma_{\rm BB}\lesssim \Gamma_{\rm vac}$, and $T_{\rm BB}$ in which
$\Gamma_{\rm BB}\gg \Gamma_{\rm vac}$, and consider either class to be
roughly equally likely.  Then we update based on the available evidence.         

To be concrete, we focus on a single question of conditional 
probability:  given being an observer like any of us living on a planet 
like ours, what is the probability that we find ourselves in a galaxy, 
surrounded by other galaxies?  Because it is far more likely for 
quantum transitions to form a BB that, by virtue of false memories, 
believes it is an observer like any of us living on a planet like ours, 
than to actually create the observer and the planetary environment, we 
should phrase the question in terms of observers holding certain 
beliefs.  We denote the belief in being an observer like any of us 
living on a planet like ours $D_\odot$, and that along with the 
additional belief of living in a galaxy surrounded by other galaxies 
$D_\oplus$.  Since we believe $D_\oplus$, the relative posteriori 
probability of theories $T_{\rm NO}$ and $T_{\rm BB}$ is
\beq
\frac{P(T_{\rm NO}|D_\oplus)}{P(T_{\rm BB}|D_\oplus)} = 
\frac{P(D_\oplus|D_\odot,T_{\rm NO})\,P(D_\odot|T_{\rm NO})\,P(T_{\rm NO})}
{P(D_\oplus|D_\odot,T_{\rm BB})\,P(D_\odot|T_{\rm BB})\,P(T_{\rm BB})} \,.
\eeq
Depending on perspective, we can take $P(D_\odot|T_{\rm NO,BB})$ to be 
unity (reflecting the fact that theory $T_{\rm NO,BB}$ predicts with 
certainty the existence of observers with belief $D_\odot$) or combine 
it with $P(T_{\rm NO,BB})$ as part of what we mean by the prior 
probability.  In either case the last two factors in the numerator are 
roughly equal to the last two factors in the denominator, by hypothesis.  
Meanwhile, theory $T_{\rm NO}$ predicts that the number of observers 
who believe $D_\oplus$ is roughly the same as the number of observers 
who believe $D_\odot$, since this theory is dominated by NOs whose 
beliefs correlate with the world around them, and whose solar systems 
typically arise as part of a galaxy surrounded by other galaxies.  
Therefore $P(D_\oplus|D_\odot,T_{\rm NO})\sim{\cal O}(1)$.  On the 
other hand, theory $T_{\rm BB}$ predicts that the number of observers 
who believe $D_\oplus$ is much smaller than the number of observers 
who believe $D_\odot$.  In the case of BBs whose beliefs correlate
with the world around them, this is because a quantum transition is
far less likely to create a large number of galaxies than to create a 
single solar system, due to the difference in mass.  In the case of 
BBs with false memories, this is because a quantum transition is far 
less likely to create beliefs $D_\oplus$ than to create beliefs 
$D_\odot$, due to the difference in entropy (information).  
Therefore $P(D_\oplus|D_\odot,T_{\rm BB})$ is exponentially suppressed 
\cite{DeSimone:2008if}, and we conclude that theory $T_{\rm NO}$ is 
far more probable than theory $T_{\rm BB}$.
   
In the context of the measure problem, one usually compares theories 
with the same local physics (including vacuum decay rates and BB 
nucleation rates), but with different measures.  Then, in analogy to
the above argument, measures that predict that BBs vastly outnumber 
NOs are deemed far less probable than measures that do not, all else 
being roughly equal.  This line of reasoning argues against the 
pocket-based measure \cite{Garriga:2005av}, or any other measure
that respects the internal symmetries of a CDL bubble by giving equal
weight to all comoving volumes at all times in a given bubble.  
(Indeed, the diverging number of BBs in a fixed comoving volume in a 
dS bubble as FRW time runs to infinity is a symptom of the fact that 
such a measure prescription does not completely regulate the 
diverging spacetime volume.  Note that such measures also suffer 
from runaway inflation.)  On the other hand, avoiding BB domination
is a non-trivial matter for any measure that does not face the other
phenomenological pathologies.  In the case of the causal patch, fat
geodesic, scale-factor, lightcone time, and CAH+ measures described 
below, it requires, roughly speaking, that the decay rate of any 
vacuum in which BBs are produced be larger than the BB formation rate 
in that vacuum, in some cases in addition to other criteria 
\cite{Bousso:2006xc,DeSimone:2008if,Bousso:2008hz}.

\subsubsection{``End of time''}

Both the global and the local approaches to the measure problem 
feature an unsettling relationship between the probabilistic outcomes 
of experiments and the times taken to perform the experiments 
\cite{Bousso:2010yn,Guth:2011ie}.  While this feature is not a paradox 
in the traditional sense---measures can have this effect while at the 
same time being logically consistent and in agreement with all 
available data \cite{Guth:2011ie}---we discuss it here in the interest 
of completeness.

Consider two sequences of events, $S_{0\to 1}$ and $S_{0\to 2}$, 
with the interval between the first and last event being $\tau_1$ and 
$\tau_2$ (with $\tau_1<\tau_2$) respectively.  To help visualize, 
consider again Figure \ref{fig:multiverse}, and imagine that each of 
these sequences of events has the same, uniform probability per unit 
three-volume to begin on the future lightcone $\Sigma_{\rm n}$ of 
the nucleation event of the largest CDL bubble displayed there.  
Also, imagine that we are in the process of computing the relative 
number of these sequences, and are correspondingly looking at 
regions below the cutoff $\Sigma_t$ or in the past lightcone of 
$\gamma$, as indicated in the diagram.  Even though the starting 
points of each sequence are given the same distribution across 
$\Sigma_{\rm n}$, since the sequence $S_{0\to 1}$ takes less time 
than the sequence $S_{0\to 2}$, more complete sequences $S_{0\to 1}$ 
can be fit under the cutoff.  This conclusion is unchanged when
we consider sequences within an ensemble of different bubbles, 
across an ensemble of different semi-classical realizations of 
spacetime.  Therefore, these measures assign higher probability
to the sequence $S_{0\to 1}$ than to $S_{0\to 2}$.

While it does not seem there is anything logically inconsistent
in this result, it can be phrased in terms of fanciful thought
experiments that lead to unsettling conclusions.  (In particular,
insistence on evolving local initial data according to unitary 
operators seems to imply one must insert an ``end of time'' to 
reproduce the predictions of these measures \cite{Bousso:2010yn}.)  
For example, the first event in each sequence could be the flipping 
of a fair coin, which determines how long a subject will be put to 
sleep.  Right after the coin is flipped, the subject should predict 
equal likelihood for heads/tails, reflecting the uniform 
distribution of starting points of the sequences $S_{0\to 1}$ and 
$S_{0\to 2}$.  However if the subject is unaware of the outcome of 
the coin flip, and is asked upon waking what was the outcome, the 
subject should predict higher likelihood for the outcome that 
results in being put to sleep for less time, reflecting the greater 
number of sequences $S_{0\to 1}$ than $S_{0\to 2}$ under the cutoff.  
Roughly speaking, for the measures discussed below the ratio of 
probabilities is of the form $e^{H\Delta\tau}$, where 
$\Delta\tau=\tau_2-\tau_1$ is the difference in intervals of the 
two sequences, and $H$ is of order the Hubble rate over the course 
of the sequences (for more precise statements, see
\cite{Bousso:2010yn,Guth:2011ie}).  Thus, the effect is only 
significant for differences in intervals on order the Hubble
time, and in particular does not contradict the outcomes of any 
known local experiments.

\subsection{Measure proposals}
\label{ssec:measureproposals}

We now briefly describe several measure proposals.  The measures 
considered here are selected because they have been studied in detail
and are known to avoid each of the phenomenological pathologies of
Section \ref{sec:pathologies} (except the ``end of time'' issue, which
as we have discussed is not a phenomenological pathology in the sense
that it does not lead to conflict with any available data).  We have two 
main goals.  One is to provide a sufficiently precise definition to 
build intuition for the predictions of each measure, as well as to raise 
any known phenomenological issues.  The other is provide the 
distribution of observers like us in bubbles like ours over the open 
FRW radial coordinate $\xi$ of a CDL bubble geometry, as this is 
relevant to the expected spatial distribution of bubble collisions one 
should expect to observe.

\subsubsection{Causal patch measure}
\label{sssec:causalpatch}

As mentioned above, the causal patch measure is a local measure, which
uses the causal patch surrounding a given worldline---either the region 
within the past lightcone of the endpoint of the worldline (where it 
hits an AdS singularity), or the causal diamond, the region within the 
intersection of this past lightcone with the future lightcone of the 
origin of the worldline---to determine which events to count and which 
to ignore.

One considers all future histories of the worldline, so that the 
relative number of times the worldline encounters a bubble of vacuum
type $i$ is given by the sum over branching ratios to semi-classical
spacetimes where this happens.  Symbolically, we write this number 
\cite{Bousso:2006ev}
\beq
n_i = \delta_{i*} + \sum_{\rm paths} \beta_{ij}\beta_{jk}\dots
\beta_{l*} \,,
\eeq
where $\beta_{ij}$ is the branching ratio for a vacuum of type $j$ 
to transition to a vacuum of type $i$, in the vicinity of the 
worldline, the sum is over all sequences of transitions that can 
take vacuum $*$ to vacuum $i$ (in any number of steps), and the 
asterisk denotes the initial vacuum in which the worldline begins.  
The branching ratios can be written 
$\beta_{ij}=\kappa_{ij}/\sum_k\kappa_{kj}$, where
\beq
\kappa_{ij}\equiv (4\pi/3)H_j^{-4}\Gamma_{ij} \,,
\eeq
with $\Gamma_{ij}$ being the decay rate (per unit four-volume) of 
vacuum $j$ to vacuum $i$.  It is common to refer to the ``prior''
probabilities among vacua in the multiverse, which reflects the
probability assigned to vacua before conditioning effects are 
included.  In the case of the causal patch measure, the prior
for vacuum $i$ is given by
\beq
{\cal I}_i = \frac{n_i}{\sum_j n_j} \,.
\label{Iprior}
\eeq     

After some reasonable approximations, it is not difficult to include
conditioning effects in the causal patch measure.  If we focus on the 
predictions of observers who reside in large gravitationally-collapsed 
structures, then we focus on vacua with a very small cosmological 
constant $\Lambda$.  Transitions to vacua with larger vacuum energies 
are suppressed by a factor $e^{-3/G\Lambda}$ relative to transitions 
to smaller vacuum energies \cite{Lee:1987qc}.  Insofar as $\Lambda$ is 
unusually small, it is unlikely that any of the direct, fastest decay 
channels connect to a dS vacuum with smaller $\Lambda$, and so one 
expects the decay rate to be dominated by transitions to AdS vacua.  
This implies that bubbles of such small-$\Lambda$ vacua are typically 
situated at the future tip of the causal diamond, right before the 
worldline encounters an AdS bubble / singularity.  Furthermore, the 
past lightcone emanating (backward) from the future tip of the casual 
patch will typically closely resemble the past lightcone of an eternal 
worldline in the bubble, because typical decays will be to vacua with 
order Planck-scale AdS curvature radii.  To be concrete, we focus on 
the case where the causal patch is the past lightcone of the endpoint 
of the worldline.  (The past lightcone and the causal diamond of a 
worldline frequently enclose the same post-inflationary spacetime 
volume in bubbles near the future tip of the causal patch, implying 
that their predictions for such regions would approximately be the 
same, however we have not performed a quantitative study of when this 
is and is not the case.) 

Accordingly, one usually writes the relative number of events of type 
$i$
\beq
{\cal N}_i \,\propto\, \sum_j\,\, {\cal I}_j\! \int_0^\infty d\tau\, 
V_{\diamond|j}(\tau)\,\rho_{i|j}(\tau)  \,,
\label{causalpatchP}
\eeq
where the number of events of type $i$ in a typical bubble of type $j$
is estimated by integrating over time the physical volume 
$V_{\diamond|j}$ enclosed by the past lightcone of an eternal worldline 
in vacuum $j$, multiplied by the four-density $\rho_{i|j}$ of events 
of type $i$ in vacuum $j$.  Each vacuum $j$ is weighted by the prior
probability ${\cal I}_j$ of this measure, and then one sums over vacua
$j$.  In AdS vacua the upper limit of integration should be the maximum 
FRW proper time reached in such bubbles.  (We suppress the dependence 
on the initial conditions.)  

Note that this probability distribution raises a few concerns.  In 
\cite{Salem:2009eh} it was shown that the factor $V_{\diamond}$ is 
much larger for observers like us in bubbles like ours but with 
negative cosmological constant, indicating a strong preference for 
observers otherwise like us to measure $\Lambda<0$.  A more careful 
consideration of very small vacuum energies has revealed a far more 
severe issue:  the distribution of $\Lambda$ grows like 
$|\Lambda|^{-2}$ for small negative $\Lambda$, and like 
$\Lambda^{-1}$ for small positive $\Lambda$ \cite{Bousso:2010zi,
Bousso:2010im}.  It seems the situation can only be resolved if we 
presume the correct measure takes a different form when describing 
AdS bubbles, and if the ultimately discrete distribution of vacuum 
energies allows for a consistent explanation for the value of 
$\Lambda$ we observe.

The expression (\ref{causalpatchP}) assumes that the events $i$ are 
distributed according to an FRW symmetry within the bubble, and  
correspondingly ignores the FRW radial coordinate $\xi$ in the bubble.  
As explained in the main text, the distribution of the effects of 
bubble collisions on the surface of last scattering generically 
depends on this coordinate, so we would like to know the distribution 
of observers like us as a function of their position $\xi$ in bubbles 
like ours.  

The worldline defining the causal patch measure could intersect our
bubble at any point along the bubble wall, depending on where our 
bubble nucleates in relation to the worldline.  The nucleation of our 
bubble is a local, random phenomenon, and therefore across the ensemble
of semi-classical future histories surrounding the worldline, the point 
of nucleation is uniformly distributed with respect to the comoving 
frame established by the initial conditions.  From the perspective of 
our bubble, the ensemble describes a congruence of comoving worldlines, 
uniformly distributed in the comoving frame of the initial conditions.  

We first take the spatially-flat chart to provide the comoving frame 
set by the initial conditions, which is appropriate for bubble 
nucleations that are far from the the initial hypersurface.  Then the 
distribution of worldlines is simply $dP\propto \Hp^3\,r^2\,dr$, which 
can be expressed in terms of the bubble coordinate $\xi$ via (\ref{r2xi}).  
If we take all observers to sit precisely on the worldline defining the 
measure, this gives
\beq
dP(\xi) \propto \frac{(1-e^{-\xi})^2\, e^{-\xi}}
{(\Hp/\Hd+ e^{-\xi})^4}\,d\xi \,.
\label{CP}
\eeq
We may also consider the closed dS chart to provide the comoving frame 
set by the initial conditions.  This would be appropriate in the
``tunneling from nothing'' scenario mentioned in the main text.  
Then the distribution of worldlines is $dP\propto\sin^2(R)\,dR$, 
which can be expressed in terms of the bubble coordinate $\xi$ via 
(\ref{R2xi}).  The resulting distribution is a complicated function
of $T_{\rm n}$, $\Hp/\Hd$, and $\xi$, yet with the aid of a computer
it can be seen to exhibit the same qualitative features as (\ref{CP}): 
the distribution falls off quadratically with $\xi$ for $\xi\ll 1$, 
and exponentially with $\xi$ for $\xi\gg 1$.  Indeed, for all positive
$T_{\rm n}$ the latter distribution is more sharply peaked at smaller
values of $\xi$ than (\ref{CP}).  (As expected, the two distributions
converge in the limit where the bubble nucleates far from the initial
hypersurface, $T_{\rm n}\to\pi/2$.)

The distribution (\ref{CP}) refers to observers who sit precisely on 
the worldline defining the measure, however the causal patch measure
counts not just these observers but all observers contained in the 
volume $V_{\diamond}$ surrounding this worldline.  Therefore, technically, 
we should convolute the distribution (\ref{CP}) over this volume, with
$V_{\diamond}$ evaluated at the time of the observer, $\tobs$.  
Nevertheless, for observers like us (who arise after inflation) in 
bubbles like ours (where cosmological constant-domination occurs before 
there can be spatial-curvature domination), it can be shown that 
$V_{\diamond}(\tobs)$ covers a comoving volume that is smaller than the 
curvature radius; therefore such a convolution does not qualitatively 
change the distribution (\ref{CP}).

\subsubsection{Fat geodesic measure}

The fat geodesic measure \cite{Bousso:2008hz} is defined in the same 
manner as the causal patch measure, but instead of counting only events 
within the causal patch of a given worldline, it counts only events 
within a (small) fixed physical distance orthogonal to the worldline.
(This should also provide a good approximation of the semi-classical 
approximation of the measure proposed in \cite{Nomura:2011dt}, modulo
a theory of initial conditions.)  As with the causal patch measure, at 
the formal level one should ignore stable Minkowski vacua, because the 
fat geodesic would enclose a four-volume that diverges with proper 
time in the bubble.  (At the practical level, however, we expect the 
number of events of any type within the fat geodesic to be finite in 
Minkowski bubbles, due to dilution in the expanding FRW background.)

Clearly, the ``prior'' probability the fat geodesic measure assigns to 
a given vacuum is the same as that assigned by the causal patch measure, 
that is (\ref{Iprior}) above.

Proceeding in analogy to with the causal patch measure, one might 
suppose the relative number of events of type $i$ can be written the 
same as (\ref{causalpatchP}), but with the volume $V_{\diamond|j}(\tau)$ 
of the causal patch in vacuum $j$ replaced by the fixed physical volume 
orthogonal to the fat geodesic (which would then factor out of the 
expression).  Taking the worldline prescription more seriously, however, 
it makes a difference whether the ``thickness'' of the fat geodesic is 
very large or not when compared to the typical separation between 
gravitationally-collapsed objects.  Taking this thickness to be very 
small (compared to any scale of interest), the fat geodesic will tend to 
fall into gravitationally bound structures, creating some bias for events 
in those regions.  (This effect is unimportant in the causal patch 
measure, because $V_{\diamond|j}$ is very large compared to the typical 
separation between gravitationally collapsed structures.)  Usually one
is interested in precisely such events, for example the measurements of 
observers residing in large galaxies.  We account for this by noting 
that the fraction of comoving worldlines that become bound by 
gravitationally-collapsed structures with mass greater than $M$, is 
precisely the corresponding collapse fraction of 
non-relativistic matter, $F_{\rm c}(M)$.  Moreover, the rate at which
worldlines are captured by structures with mass equal to $M$ is 
$dF_{\rm c}(M)/d\tau$.  Thus, the relative 
number of these events $i$ can be written
\beq
{\cal N}_i \,\propto\, \sum_j\,\, {\cal I}_j\! \int_0^\infty d\tau\,
\frac{dF_{{\rm c}|j}(M,\tau)}{d\tau}\, \tilde{\rho}_{i|j}(M,\tau)  \,,
\label{fatgeodesicP1}
\eeq 
where $\tilde{\rho}_{i|j}(M,\tau)$ is the density of events $i$ in 
collapsed structures that have mass $M$ at time $\tau$.  In the 
absence of gravitational collapse, the naive expectation is correct, 
and we have
\beq
{\cal N}_i \,\propto\, \sum_j\,\, {\cal I}_j\! \int_0^\infty d\tau\,
\rho_{i|j}(\tau)  \,.
\label{fatgeodesicP2}
\eeq          

The fat geodesic measure has one phenomenological issue that might
raise concern \cite{Bousso:2010im}.  The measure is defined to 
count all events within a fixed physical distance of a given 
worldline, but in AdS vacua the scale factor goes to zero at the 
AdS singularity, meaning the fat geodesic grows to enclose a 
diverging comoving volume, and likewise counts a diverging quantity 
of matter there.  It is unclear how seriously to take this effect, 
since it seems reasonable to suppose that the crunching bubble 
destroys all observers long before the singularity.  Still, it is 
possible that among a class of observers among whom we should
consider ourselves typical, the vast majority arise after 
scale-factor turnaround in an AdS bubble.

The above distributions assume the events $i$ are uncorrelated with
the open FRW radial coordinate $\xi$, which has been ignored.  The
distribution of observers like us as a function of this coordinate
can be computed in analogy to in Section \ref{sssec:causalpatch} 
(the bias resulting from the tendency of the worldline to gravitate 
toward bound structures is on average independent of $\xi$).  In the
case of the fat geodesic measure, convolution over the volume of the
causal region $V_{\diamond|j}(\tobs)$ should be replaced by 
convolution over the fixed volume contained within the fat geodesic.  
The latter is taken to be very small, and therefore does not affect 
the distribution.  Therefore, assuming the fat geodesic is comoving
in the spatially-flat dS frame, we have  
\beq
dP(\xi) \propto 
\frac{(1-e^{-\xi})^2\, e^{-\xi}}
{(\Hp/\Hd+ e^{-\xi})^4}\,d\xi \,.
\label{FG}
\eeq  
As described in Section \ref{sssec:causalpatch}, the distribution 
when the fat geodesic is comoving in the closed dS frame is 
qualitatively similar, but more sharply peaked at a somewhat smaller
value of $\xi$.

\subsubsection{Scale-factor cutoff measure}
\label{sssec:sfc}

As a global measure, the scale-factor cutoff begins with a finite, 
spacelike hypersurface $\Sigma_0$, on which the ``scale-factor 
time'' $t$ is set to zero, and studies the evolution of $t$ along
a congruence of geodesics orthogonal to $\Sigma_0$.  The scale-factor 
time is defined according to
\beq
dt = H d\tau \,,
\label{sfdef}
\eeq
where $H$ is some measure of the expansion along the congruence and 
$\tau$ is the proper time.  To be concrete, we choose 
$H = (1/3)\,u^\mu_{\phantom{\mu};\,\mu}$, i.e.~we take $H$ to be a 
third of the divergence of the four-velocity field along the 
congruence.  (At caustics one uses the smallest scale-factor time.)  
The scale-factor cutoff measure counts only those events that occur
between $\Sigma_0$ and a constant scale-factor hypersurface 
$\Sigma_t$, in the end taking $t\to\infty$.  As we have defined it, 
the cutoff hypersurface $\Sigma_t$ develops a very complicated 
intrinsic geometry in the vicinity of gravitationally-bound 
structures.  One might expect the ``correct'' measure to be more 
smooth over such scales.  As a technical matter this can be 
accomplished by augmenting the cutoff hypersurface with the future 
lightcones of all points on $\Sigma_t$ \cite{DeSimone:2008if}.  This 
also gives a prescription for how to deal with AdS bubbles.  Outside 
of AdS bubbles, the resulting cutoff $\Sigma_+$ closely resembles 
the cutoff that results from defining $H$ to be the FRW Hubble rate 
\cite{DeSimone:2008bq} (this is because the light-crossing time 
between voids is much smaller than the Hubble time), but has the 
advantage of being simple to define in geometries that lack any of 
the FRW symmetries.  The cutoff in AdS bubbles is described below.  

Because spacetime volume grows exponentially with scale-factor time, 
most of the four-volume between the hypersurfaces $\Sigma_0$ and 
$\Sigma_t$ is near $\Sigma_t$.  Moreover, the physical three-volume 
$V_i$ of a given dS vacuum $i$ on $\Sigma_t$ approaches an attractor 
with increasing $t$.  This can be seen by studying the coarse-grain 
rate equation \cite{Garriga:2005av,DeSimone:2008if},   
\beq
\frac{dV_i}{dt} = 3V_i+\sum_j \kappa_{ij}V_j - \sum_j\kappa_{ji}V_i \,.
\label{rate}
\eeq
Coarse-graining in this context means treating a bubble nucleation as 
an instantaneous swap of volume $(4\pi/3)H_j^{-3}$ of the decaying 
vacuum $j$ for the same volume of vacuum $i$, and ignoring any 
transitory evolution between bubble nucleation and false-vacuum energy 
domination.  The subsequent change in volume due to expanding bubble 
walls is negligible next to the growth in volume due to scale-factor 
expansion, represented by the first term in (\ref{rate}).   

The solution of (\ref{rate}) can be written (for a dS vacuum $i$) 
\cite{Garriga:2005av,DeSimone:2008if}
\beq
V_i(t) \propto s_i\, e^{(3-q)t} + \ldots \,,
\label{ratesol}
\eeq
where $s_i$ is the eigenvector of the matrix 
$\kappa_{ij}-\delta_{ij}\sum_k \kappa_{ki}$ with the 
smallest-magnitude eigenvalue $-q$ (it happens that $q$ is always 
pure-real and positive), and the ellipses denote terms that grow more
slowly than $e^{(3-q)t}$.  In the simplest case, one finds one component
of $s_i$ to be much larger than the rest; it corresponds to the 
slowest-decaying dS vacuum, called the dominant vacuum, which has a 
total decay rate given by $q$, up to corrections proportional to the 
``upward'' tunneling rates to states with larger vacuum energies.  
(More complicated scenarios include when an isolated set of vacua 
collectively behave as the dominant vacuum, see \cite{DeSimone:2008if}.)  
The other components of $s_j$ are suppressed relative to the dominant 
one by factors including the upward tunneling rate from the dominant 
vacuum.  

Going beyond the coarse-grain distribution of volume fractions, the 
relative number of events of type $i$ can be approximated as
\beq
{\cal N}_i \propto \sum_{j,k} \kappa_{jk}s_k \int_0^{t_{\rm c}} 
dt_{\rm n}\, e^{(3-q)t_{\rm n}} \int_0^{\tau_{\rm c}} 
d\tau\,a^3(\tau)\,\rho_{i|j}(\tau) 
\int_0^{\xi_{\rm c}} d\xi\,\sinh^2(\xi) \,, 
\label{sfp}
\eeq 
which we describe as follows.  The two integrations to the right
count the number of events of type $i$ in a bubble of vacuum $j$ 
below the cutoff $\Sigma_+$, this cutoff determining the limits of 
integration $\tau_{\rm c}$ and $\xi_{\rm c}$.  The other integral 
sums over bubble nucleation times $t_{\rm n}$ (up to the cutoff 
time $t_{\rm c}$), while the factors out front account for the 
probability to nucleate a vacuum of type $j$ in a parent of vacuum 
type $k$, times the volume fraction in vacuum $k$, summed over all 
vacua $j$ and $k$.  The factor $e^{-qt_{\rm n}}$ comes from 
(\ref{ratesol}); however next to the volume expansion factor 
$e^{3t_{\rm n}}$ it can be ignored.

The limits of integration $\tau_{\rm c}$ and $\xi_{\rm c}$ are 
determined as follows.  If the bubble of vacuum $j$ nucleates more 
than a few Hubble times after its parent of vacuum $k$, then 
the congruence that defines the measure will be comoving in the 
spatially-flat frame.  For simplicity we use the spatially-flat
frame to provide a qualitative picture of the general case.  In dS 
bubbles the hypersurface $\Sigma_+$ can then be approximated by 
(\ref{sftime}), substituting $\Delta t_{\rm sf}\to t_{\rm c}-t_{\rm n}$ 
and $\Hd\tau_{\rm ref}\to\ln[2\Hd a(\tau)]$.  Solving this equation
for $\xi$ gives $\xi_{\rm c}(t_{\rm c},t_{\rm n},\tau)$; solving
for $\tau$ with $\xi=0$ gives $\tau_{\rm c}(t_{\rm c},t_{\rm n})$.
In AdS bubbles, $\xi_{\rm c}(\tau)$ can become timelike---label 
the earliest point at which this happens $(\tau_*,\xi_*)$---then 
one should use $\xi_{\rm c}=\xi_* - \int_{\tau_*}^\tau d\tau'/a(\tau')
=\xi_*-\eta_*(\tau)$ when $\xi\leq\xi_*$ (with $\tau_{\rm c}$ 
determined by solving for $\xi_{\rm c}=0$), otherwise proceeding 
as before.  

Notice that in dS bubbles, we can write $\xi_{\rm c} = g(x)$, 
where $x=t_{\rm c}-t_{\rm n}-\ln[\Hd a(\tau)]$.  Performing the 
integral over $\xi$ in (\ref{sfp}), reversing the order of 
integration between $t_{\rm n}$ and $\tau$, exchanging $t_{\rm n}$ 
for $x$, and finally integrating over $x$ (noting the limit 
$t_{\rm c}\to\infty$), we obtain
\beq
{\cal N}_i \propto \sum_{j,k} \kappa_{jk}s_k \int_0^\infty 
d\tau\,\rho_{i|j}(\tau) \,, 
\label{sfp2}
\eeq
where we have dropped a $\tau$-independent dimensionless integral
involving the ratio $\Hd/\Hp$.  One can perform the same 
manipulations in the case of AdS bubbles, but the lightcone 
prescription described above prevents the anticipated 
simplifications.  An alternative, qualitative approximation is to 
use the lightcone prescription to approximate $H$ using the FRW 
Hubble rate in the bubble, but instead of using the cutoff 
$\xi_{\rm c}=\xi_*-\eta_*(\tau)$ when $\xi\leq\xi_*$, simply 
ignore all events that occur after scale-factor turnaround at 
time $\tau_{\rm turn}$.  Then we find
\beq
{\cal N}_i \propto \sum_{j,k} \kappa_{jk}s_k \int_0^{\tau_{\rm turn}} 
d\tau\,\rho_{i|j}(\tau) \,, 
\label{sfp3}
\eeq
where we have dropped the same dimensionless integral as in
(\ref{sfp2}).  These distributions are studied in 
\cite{DeSimone:2008bq,DeSimone:2009dq} and do not appear to have 
any phenomenological shortcomings.

We can also use (\ref{sfp}) to compute the distribution of observers
like us over the radial FRW coordinate $\xi$.  Conditioning on our
observation of the present FRW time $\tobs$, we write
$\rho_{i|j}(\tau)\propto \delta(\tau-\tobs)$.  Reversing the order
of integration between $\tau$ and $\xi$, and then reversing the order
of integration between $\xi$ and $t_{\rm n}$, integrating over 
$\tau$ and $t_{\rm n}$, and then taking $dP(\xi)$ as the argument of 
the remaining integral over $\xi$, we obtain
\beq
dP(\xi) \propto 
\frac{(1-e^{-\xi})^2\, e^{-\xi}}
{(\Hp/\Hd+ e^{-\xi})^4}\,d\xi \,.
\eeq  
Interestingly, it is the same distribution as for the local measures 
computed above, when they are defined with respect to a geodesic that
is comoving in the spatially-flat frame.

\subsubsection{Global time cutoffs and bulk/boundary duality}

Motivated by the hope that gravitational physics in ``bulk'' 
spacetime might be dual to a (Euclidian) conformal theory on the 
boundary at future infinity, there has been recent interest in 
understanding what (if any) bulk global time cutoff might correspond 
to an ultra-violet (UV) cutoff on the boundary theory.  This approach
to the measure problem was first described in 
\cite{Garriga:2008ks,Garriga:2009hy}, 
where it was suggested that a UV cutoff on the boundary theory might 
correspond to a scale-factor cutoff in the bulk.  However, it was 
later argued in \cite{Bousso:2009dm,Bousso:2010id} that a global 
``lightcone time'' cutoff better captured the spirit of AdS/CFT 
correspondence.  Roughly speaking, the lightcone time at a given 
point in spacetime corresponds to some measure of the size of the 
future lightcone of the point at future infinity.  (Defining the 
metric over points at future infinity, including how to address 
AdS and Minkowksi bubbles, is a crucial outstanding problem; 
\cite{Garriga:2008ks} and \cite{Bousso:2010id} suggest two different
proposals.)  As a global time cutoff, the lightcone time measure 
features an attractor solution to the volume-fraction rate equations 
in a coarse-grained description of the multiverse (as described for 
the scale-factor cutoff measure above), yet its predictions are 
otherwise the same as for the causal patch measure, after choosing 
suitable initial conditions for the latter \cite{Bousso:2009mw}.  
Therefore, we do not comment further on this measure, ascribing to 
it the results of Section \ref{sssec:causalpatch}.  

It has also been suggested that a UV cutoff on the boundary theory 
corresponds to a (lower) cutoff on the size of the comoving apparent 
horizon (CAH) in the bulk, with ``comoving'' coordinates defined 
with respect to a congruence orthogonal to some initial spacelike 
hypersurface $\Sigma_0$ \cite{Vilenkin:2011yx}.  (With the same 
approach to constructing a measure over points at future infinity,
the lightcone time cutoff could be viewed as a comoving horizon
cutoff in the bulk.)  For practical purposes, the ``CAH time'' can 
be defined 
\beq
\theta = \dot{a} \,,
\eeq     
where $a$ is the expansion factor along the congruence emanating
from $\Sigma_0$, and the dot denotes differentiation with respect 
to proper time (see \cite{Vilenkin:2011yx} for a more general
discussion).  As it is here defined, the CAH time can halt and even
decrease as a function of proper time along a given element of
the defining geodesic congruence.  This can be dealt with in the
same way that the same underlying issue was handled with respect to
the scale-factor cutoff measure:  by augmenting the cutoff 
hypersurface $\Sigma_\theta$ with the future lightcones of all 
points on $\Sigma_\theta$ \cite{Vilenkin:2011yx}.  The resulting 
measure has been called CAH+, and its phenomenology is studied in 
\cite{Salem:2011mj}.

In the context of the coarse-grained picture described for the 
scale-factor cutoff in Section \ref{sssec:sfc}, the rate equation 
(\ref{rate}) is modified by the variable transformation 
$t\to\ln(\theta/\theta_0)$ in switching from scale-factor time to 
CAH time, and a factor $H_j^3/H_i^3$ that appears within the first
sum on the right-hand size of (\ref{rate}) \cite{Salem:2011mj}.  Here 
$\theta_0$ is a ``universal'' scale that must be introduced; it can 
be taken to correspond to the CAH time on the initial hypersurface 
$\Sigma_0$.  The solution is simply related to (\ref{ratesol}); for 
a dS vacuum $i$
\beq
V_i(t) \propto H_i^{-3}\, s_i\, (\theta/\theta_0)^{3-q} + \ldots \,,
\eeq
where the quantities $s_i$ and $q$ are all the same 
as for the scale-factor cutoff, and the ellipses denote 
terms that do not grow as quickly as $\theta^{3-q}$.    

Going beyond the coarse-grained description, in analogy to with the
scale-factor cutoff measure the relative number of events of type 
$i$ can be approximated
\beq
{\cal N}_i \propto \sum_{j,k} \kappa_{jk}s_k 
\int_{\theta_0}^{\theta_{\rm c}} 
d\ln(\theta_{\rm n}/\theta_0)\, \theta_{\rm n}^{3-q} 
\int_0^{\tau_{\rm c}} d\tau\,a^3(\tau)\,\rho_{i|j}(\tau) 
\int_0^{\xi_{\rm c}} d\xi\,\sinh^2(\xi) \,. 
\label{cah}
\eeq 
As before, the power of $q$ can be ignored.  The limits of 
integration $\tau_{\rm c}$ and $\xi_{\rm c}$ are determined as 
follows.  As with the the scale-factor cutoff measure, we take
the comoving frame set by the initial hypersurface to correspond
to the spatially-flat dS frame in the vicinity of the bubble.  
The corresponding CAH cutoff inside the bubble is calculated in 
\cite{Salem:2011mj}, and gives
\beq
\theta_{\rm c} = \theta_{\rm n} 
\left[\frac{(1+e^{-\xi_{\rm c}})^2(\Hp+\Hd e^{-\xi_{\rm c}})^4}
{32\Hp^3(\Hp+\Hd)}\right]^{1/3}\! 
\frac{\dot{a}(\tau)}{\dot{a}(\tau_{\rm ref})}\, 
e^{\xi_{\rm c}+\Hd\tau_{\rm ref}} \,,
\eeq   
where $\tau_{\rm ref}$ is some reference time deep within the 
inflationary epoch.  Inverting this equation gives
$\xi_{\rm c}(\theta_{\rm c},\theta_{\rm n},\tau)$; solving
for $\tau$ with $\xi_{\rm c}=0$ gives 
$\tau_{\rm c}(\theta_{\rm c},\theta_{\rm n})$.  The resulting
cutoff hypersurface becomes null near the onset of reheating---label
the point at which this happens $(\tau_*,\xi_*)$---in this case one 
should use $\xi_{\rm c}=\xi_* - \int_{\tau_*}^\tau d\tau'/a(\tau')
=\xi_*-\eta_*(\tau)$ when $\xi\leq\xi_*$ (with $\tau_{\rm c}$ 
determined by solving for $\xi_{\rm c}=0$), unless the 
$\theta=\theta_{\rm c}$ hypersurface provides a stronger cutoff.  
In AdS and Minkowksi bubbles, the above null surface defines the 
cutoff at all times after reheating.  In dS bubbles, for sufficiently 
large $\theta_{\rm c}/\theta_{\rm n}$ and at sufficiently late FRW 
times in the bubble (long after cosmological-constant domination in 
bubbles like ours), the $\theta=\theta_{\rm c}$ hypersurface can 
provide a stronger cutoff than the above null surface.  

The expression (\ref{cah}) can be manipulated as was done for 
the corresponding expression in the scale-factor cutoff measure.  
The details are discussed in \cite{Salem:2011mj}, and give
\beq
{\cal N}_i \propto \sum_{j,k} \kappa_{jk}s_k \int_0^\infty 
d\tau\,\rho_{i|j}(\tau)\, h_j(\tau) \,. 
\label{cah2}
\eeq
In dS bubbles $h_j(\tau)$ is the smaller of $e^{-3\eta(\tau)-3\Ne}$ 
and $\big[2H_j\,a(\tau)\big]^{-3}$, where $\eta(\tau)$ is the FRW 
conformal time in the bubble (defined so that $\eta=-(\Hd a)^{-1}$ 
deep within the inflationary epoch), $\Ne$ is the number of 
inflationary $e$-folds of expansion, and consistent with the 
notation above $H_j$ is the Hubble rate associated with the 
late-time cosmological constant domination.  In AdS and Minkowksi 
bubbles, $h_j(\tau)$ is always given by $e^{-3\eta(\tau)-3\Ne}$ (and 
the upper limit of integration should be the maximum FRW proper 
time reached in such bubbles).  The above result is valid only at 
times after reheating, and we have dropped a $\tau$-independent 
dimensionless integral involving the ratio $\Hd/\Hp$.  As described 
in \cite{Salem:2011mj}, the result does not appear to have any 
phenomenological shortcomings.   

Finally, we can use (\ref{cah}) to compute the distribution of 
observers like us over the radial FRW coordinate $\xi$.  As before, 
we condition on our observation of the present FRW time $\tobs$, 
and write $\rho_{i|j}(\tau)\propto \delta(\tau-\tobs)$.  Reversing
the order of integration between $\tau$ and $\xi$, and then 
reversing the order of integration between $\xi$ and 
$\theta_{\rm n}$, integrating over $\tau$ and $\theta_{\rm n}$, 
and then taking $dP(\xi)$ as the argument of the remaining integral 
over $\xi$, we obtain 
\beq
dP(\xi) \propto \frac{(1-e^{-2\xi})^2\,e^{-\xi}}
{(1+e^{-\xi-\Delta\eta})^2(\Hp/\Hd+e^{-\xi-\Delta\eta})^4}\,d\xi \,,
\eeq
where $\Delta\eta=\eta(\tobs)-\eta(\tau_\star)$ is the change in 
FRW conformal time between reheating and the time of observers.  
The distribution differs from those above only in the appearance 
of $\Delta\eta$ in some of the exponents, which arise as an effect 
of the null surface cutoff.  Because $\Delta\eta\ll 1$ (see Section 
\ref{ssec:modelparameters}) this is not a qualitatively 
significant difference.

%\bibliographystyle{JHEP}
%\bibliography{ETERNAL}{}

\providecommand{\href}[2]{#2}\begingroup\raggedright\endgroup

\end{document}